\documentclass[sigconf,screen]{acmart}

\pdfoutput=1

\usepackage{amsfonts}
\usepackage{url}
\usepackage[ruled,vlined]{algorithm2e}
\usepackage{multirow}
\usepackage{nicefrac}
\usepackage{graphicx}
\SetProcNameSty{textsc}
\usepackage{listings}
\lstdefinelanguage{futhark}
{
  morekeywords={
    do,
    else,
    for,
    fun,
    if,
    in,
    include,
    let,
    def,
    loop,
    struct,
    then,
    type,
    val,
    while,
    with,
    module,
    where,
    sort,
    multired,
    reverse,
    zip3,
    unzip3,
    copy,
    gather,
    withacc,
    scratch,
    sum
  },
  sensitive=true, 
  morecomment=[l]{--}, 
  morecomment=[s]{\{-}{-\}}, 
  morestring=[b]", 
  literate={\\}{\fn}{1} {->}{$\rightarrow$}{1} {<-}{$\leftarrow$}{1} {|>}{$\pipe$}{1},
}

\lstdefinelanguage{corefuthark}
{
  morekeywords={
    do,
    else,
    for,
    fun,
    if,
    in,
    include,
    let,
    loop,
    struct,
    then,
    type,
    val,
    while,
    with,
    module,
    where,
  },
  sensitive=true, 
  literate={\\}{\fn}{1} {->}{$\rightarrow$}{1} {<-}{$\leftarrow$}{1},
  moredelim=**[is][\color{red}]{@}{@},
  morecomment=[l]{--}, 
  morecomment=[s]{\{-}{-\}}, 
  morestring=[b]" 
}

\usepackage{xcolor}
\definecolor{eclipseBlue}{RGB}{42,0.0,255}
\definecolor{eclipseGreen}{RGB}{63,127,95}
\definecolor{eclipsePurple}{RGB}{127,0,85}

\newcommand{\kw}[1]{\mbox{\texttt{\bfseries{#1}}}}

\newcommand{\Map}{\kw{map}}
\newcommand{\fn}{\ensuremath{\lambda}}
\newcommand{\pipe}{\ensuremath{\triangleright}}

\newcommand{\FnU}[2]{\fn#1 \rightarrow #2}
\newcommand{\Reduce}{\kw{reduce}}

\newcommand{\Scan}{\kw{scan}}

\newcommand{\adj}[1]{\overline{#1}}
\newcommand{\Reverse}{\kw{reverse}}

\newcommand{\ttt}[1]{\texttt{#1}}
\newcommand{\Stm}[2]{\kw{let} ~ #1 ~ \ttt{=} ~ #2}

\newcommand{\Stmpb}[2]{\kw{let} ~ #1 ~ \overline{\ttt{+}}\ttt{=} ~ #2}

\newcommand{\AdOv}[1]{\frac{\text{#1}}{\text{Prim}}}

\newcommand{\altfrac}[2]{\ifmmode\def\tmp{$}\else\def\tmp{}\fi\mbox{%
    {\raisebox{.24\ht\strutbox}{\tmp#1\tmp}}%
    \kern-2.2pt\scalebox{1.6}[1.5]{/}\kern-1.8pt%
    {\tmp#2\tmp}%
}}

\newcommand{\nicediff}[2]{\smash{\altfrac{\partial #1}{\partial #2}}}

\DeclareRobustCommand{\adjoint}[1]{\smash{\overline{\texttt{#1}}}}
\newcommand{\fut}[1]{\lstinline[language=futhark,mathescape=true,basicstyle=\normalsize\ttfamily,keepspaces=true]!#1!}

\DeclareRobustCommand*{\ola}{\overleftarrow}

\theoremstyle{definition}

\begin{document}

\newcommand{\Fun}{\textsc{Fun}}
\newcommand{\FunMem}{\textsc{FunMem}}
\newcommand{\Imp}{\textsc{Imp}}

\title{Reverse-Mode AD of Reduce-by-Index and Scan in Futhark}


\author{Lotte Maria Bruun}
\email{xts194@alumni.ku.dk}
\affiliation{
  \institution{University of Copenhagen}
  \country{Denmark}}
\author{Ulrik Stuhr Larsen}
\email{usl@di.ku.dk}
\affiliation{
  \institution{University of Copenhagen}
  \country{Denmark}}
\author{Nikolaj Hinnerskov}
\email{nihi@di.ku.dk}
\affiliation{%
  \institution{University of Copenhagen}
  \country{Denmark}}
\author{Cosmin Oancea}
\email{cosmin.oancea@di.ku.dk}
\affiliation{
  \institution{University of Copenhagen}
  \country{Denmark}}

\begin{abstract}
  We present and evaluate the Futhark implementation of reverse-mode
  automatic differentiation (AD) for the basic blocks of parallel
  programming: reduce, prefix sum (scan), and reduce by index.
  We first present derivations of general-case algorithms,
  and then discuss several specializations that result in
  efficient differentiation of most cases of practical
  interest.
  We report an experiment that evaluates the performance
  of the differentiated code in the context of GPU execution,
  and highlights the impact of the proposed specializations as well as
  the strengths and weaknesses of differentiating at high level
  {\em vs.} low level (i.e., ``differentiating the memory'').
\end{abstract}

\maketitle

\section{Introduction}
\label{sec:introduction}

\enlargethispage{\baselineskip}

Nowadays, most domains have embraced machine learning (ML) methods,
which fundamentally rely on gradients to learn.
But even in the absence of ML,
the computation of gradients is essential in many
compute-hungry applications ranging over various domains,
such as risk analysis of large portfolios utilizing complex pricing
methods in finance~\cite{fin-app-ad, ad-in-finance, LexiFiPricing},
retrieval (and tuning) of parameters and associated uncertainties 
from satellite products in remote 
sensing~\cite{retrive-land-surface-params,remote-sens-satellite-aod,bfast},
solving non-linear inverse problems or for sensitivity analysis of
large numerical simulations in physics~\cite{geo-physics,ad-crop-model}.
%

Automating the computation of derivatives --- known as automatic
differentiation (AD) --- has been a central contributor to
facilitating advancements in such domains, for example in
designing and training new ML models~\cite{baydin-AD-survey}.
As such, an argument can be made that AD should be
a first-class citizen in high-level parallel
languages~\cite{10.1145/3473593,futhark-ppopp}. 
This requires algorithms that offer reliable and efficient
differentiation of (unrestricted) parallel programs that
scales well on modern, highly parallel hardware, such as GPUs. 

A feasible way of achieving this is to differentiate at a
``high level'', by building the AD algorithm around
higher-order array combinators --- common to functional
programming --- whose richer semantics allows to lift
the level of abstraction at which the compiler reasons. 
The first step in this endeavor
is to develop efficient rules for differentiating such
parallel combinators.

This paper presents and evaluates algorithms for
reverse-mode differentiation of reduce, reduce-by-index
and scan, which are implemented in (but not restricted to
the context of) the Futhark language~\cite{futhark-pldi}.
For each combinator, we present a ``general-case''
algorithm that typically re-writes the differentiation
of the combinator in terms of a less efficient combinator.
For example, reduce requires prefix sum (scan), reduce-by-index
require multi-scan, and scan's differentiation is not AD efficient.
(The algorithms for reduce and reduce-by-index are AD efficient,
but have largish constants).

These inefficiencies motivate the development of a set of
specializations that significantly reduce the AD overheads
of most cases of practical interest. Specializations include:
\begin{itemize}
\item addition, min, max and multiplication --- these are known
      and are not claimed as contributions, but are treated here
      for completeness,

\item vectorized operators, which are reduced to scalar
      operators by re-write rules that interchange the
      encompassing reduce(-by-index) or scan with the
      vectorizing map,

\item invertible commutative operators that,
      for example, allow the differentiation of
      reduce(-by-index) to be written in terms of
      a reduce(-by-index) with an extended operator,
      rather than in terms of (multi-)scan,

\item simple sparsity (compiler) optimizations that exploit
      the block-diagonal structure of Jacobians and, for
      example, allow differentiating a scan with $5\times 5$
      matrix multiplication operator at a reasonable
      $7\times$ AD overhead.
\end{itemize}

Note that the differentiation of reduce has been
(briefly) covered in \cite{futhark-ad-sc22} and is not a contribution of this
paper; we still recount it in detail for completeness and
because the rationale behind it drives the treatment of the
other operators.

Finally and most importantly, we report an experiment
that evaluates the practical GPU performance of the
reverse-mode differentiation of reduce, scan and reduce
by index on various operators. Comparisons are made with
algorithms of the closest-related approach~\cite{PPAD},
which we have implemented in Futhark.
The evaluation demonstrates significant performance gains,
and that most operators can be differentiated on GPU quite 
efficiently, while very few of them (e.g., reduce-by-index
with saturated addition) appear better suited to
lower-level approaches that ``differentiate the memory''.

In summary, key contributions of this paper are:
\begin{itemize}
\item[1] ``general-case'' reverse-mode AD algorithms for
      reduce-by-index and scan, the former of which is AD efficient,
      
\item[2] a set of specializations that offer practical
      efficiency for most cases of interest,
      
\item[3] to our knowledge, the first evaluation of the GPU
      performance of reverse-mode AD for reduce(-by-index) and
      scan; the evaluation demonstrates the claims and
      the impact of the proposed specializations
      and highlights the strengths and weaknesses of
      the approach of differentiating at a high level.      
\end{itemize}



\section{Preliminaries}
\label{sec:prelim}

This section provides a brief overview of the
functional language used to discuss the differentiation
algorithms and a brief introduction to reverse-mode
automatic differentiation (AD), which are hopefully
sufficient to understand the rest of the paper.

\subsection{Brief Overview of the Futhark Language}

Futhark is a purely-functional parallel-array language that
borrows its syntax from a combination of Haskell and ML,
and in which parallelism is explicitly expressed by means
of a nested composition of standard second-order array
combinators (SOAC), such as map, reduce, scan, and scatter
(parallel write).
Scatter has type:
\[
\kw{scatter} ~:~ \forall n,m,\alpha. \ *[n]\alpha \ \rightarrow \ [m]\kw{i64} \ \rightarrow \ [m]\alpha \ \rightarrow \ *[n]\alpha
\]
where \fut{i64} denotes the $64$-bits integral type,
$[n]\alpha$ denotes a size-typed~\cite{futhark-size-types}
array of length $n$, and $*[n]\alpha$ denotes a unique
type, e.g., when used as an argument it means that the
corresponding array is consumed by the scatter operation,
and when used for the result it means that it does not
alias any of the non-unique arguments. 

Semantically, scatter ``updates in place''
the first array argument at the indices specified in
the second array with the corresponding values stored
in the third argument. Scatter has work $O(m)$ and
depth $O(1)$. The other combinators are standard.

For brevity, the notation used in this paper is informal
and omits universal quantification and types, whenever
they are easily inferable by the reader.



\subsection{Brief Introduction to Reverse-Mode AD}

\begin{figure}[t]
\begin{minipage}[t]{.35\columnwidth}
\begin{lstlisting}[language=futhark, numbers=left]
P($x_0$,  $x_1$): 
  $t_0$ = sin($x_0$)
  $t_1$ = $x_1$ $\cdot$ $t_0$
  $y$ = $x_0$ + $t_1$
  $\kw{return}$ $y$
  
  
  
  
  
  
  
  
  
\end{lstlisting}
\end{minipage}
\begin{minipage}[t]{.63\columnwidth}
\begin{lstlisting}[language=futhark, numbers=left]
P'($x_0$,  $x_1$): 
  $t_0$ = sin($x_0$)
  $t_1$ = $x_1$ $\cdot$ $t_0$
  $y$ = $x_0$ + $t_1$
  $\adj{y}$ = 1
  $\adj{x_0}$ = 0, $\adj{t_1}$ = 0
  $\adj{x_0}$ += 1 $\cdot$ $\adj{y}$   -- $\frac{\partial(x_0+t_1)}{\partial{x_0}} \ \equiv \ 1$ 
  $\adj{t_1}$ += 1 $\cdot$ $\adj{y}$   -- $\frac{\partial(x_0+t_1)}{\partial{t_1}} \ \equiv \ 1$
  $\adj{x_1}$ = 0, $\adj{t_0}$ = 0
  $\adj{x_1}$ += $t_0$ $\cdot$ $\adj{t_1}$   -- $\frac{\partial(x_1 \cdot t_0)}{\partial{x_1}} \ \equiv \ t_0$
  $\adj{t_0}$ += $x_1$ $\cdot$ $\adj{t_1}$   -- $\frac{\partial(x_1 \cdot t_0)}{\partial{t_0}} \ \equiv \ x_1$
  $\adj{x_0}$ += cos($x_0$) $\cdot$ $\adj{t_0}$ 
                -- $\frac{\partial{\sin(x_0)}}{\partial{x_0}} \equiv \text{cos}(x_0)$
  $\kw{return}$ ($\adj{x_0}$, $\adj{x_1}$)
\end{lstlisting}
\end{minipage}\vspace{-2ex}
\caption{Simple example demonstrating reverse-mode differentiation of straight-line scalar code.}
   \label{fig:eg-revad-basic}
\end{figure}

The first-order partial derivatives of a differentiable function\\
$f : \mathbb{R}^n -> \mathbb{R}^m$ forms an $m \times n$ Jacobian matrix:
\[
\mathbf{J} =
\begin{bmatrix}
  \frac{\partial f_1}{\partial x_1} & \cdots & \frac{\partial f_1}{\partial x_n} \\
  \vdots & \ddots & \vdots \\
  \frac{\partial f_m}{\partial x_1} & \cdots & \frac{\partial f_m}{\partial x_n}
\end{bmatrix}
\]
Reverse mode AD computes a row of the Jacobian at a time.
(In contrast, forward mode AD computes a column of the Jacobian
at at time. Reverse mode AD is thus preferred when the size of
the result is much smaller than the input size.)

For a program $P(\ldots, x_i, \ldots) = y \in \mathbb{R}$,
reverse-mode AD computes the {\em adjoint} of each
(intermediate) program variable $t$, denoted 
$\overline{t} = \frac{\partial y}{\partial t}$, 
that captures the sensitivity of the result to
changes in $t$.

The {\em primal trace} (original program) is first executed
to save intermediate program values on a {\em tape} (abstraction).
The tape is subsequently used by the {\em return sweep}, which
computes the {\em adjoint} of each variable in reverse program order.

Initially $\adj{y} = \frac{\partial{y}}{\partial{y}} = 1$, and eventually
the adjoints of the input $\overline{x_i} = \frac{\partial{y}}{\partial{x_i}}$
are computed by applying the core re-write rule:

\begin{equation} \label{eq:rewrite-adj}
v \ = \ f(a,b,\ldots) \ \Longrightarrow \ 
\begin{aligned}
v=& f(a,b,\ldots) \\[-5pt]
&\hspace{-5pt}\mbox{\scriptsize$\vdots$}\\[-5pt]
\adj{a} \mathrel{+}=& \frac{\partial f(a,b,\ldots)}{\partial a} \adj{v} \\
\adj{b} \mathrel{+}=& \frac{\partial f(a,b,\ldots)}{\partial b} \adj{v} \\
\ldots 
\end{aligned}
\end{equation}
\noindent where the vertical dots correspond to the statements
of the primal trace and their differentiation
(on the return sweep). In particular,
the vertical dots compute the final value of $\adj{v}$ -- because
$v$ cannot possibly be used before its definition --- and partial
values for $\adj{a}$ and $\adj{b}$, i.e., corresponding to their
uses after the statement $v \ = \ f(a,b,\ldots)$ in the
original program.

Figure~\ref{fig:eg-revad-basic} demonstrates how reverse-mode 
AD is applied to a simple example consisting of straight-line
scalar code. The left-hand side shows the original program.
The right-hand side shows the differentiated code:
\begin{description}
\item[lines 2-4:]
    the primal trace is re-executed to bring into scope
    the values of $t_0$ and $t_1$ which are used in
    differentiation;
\item[line 5:] 
    the adjoint of the result is initialized to
    $\frac{\partial{y}}{\partial{y}} = 1$;
\item[lines 6-8] correspond to differentiating
    the last statement of the original program (line $4$)
    by the application of re-write rule~\ref{eq:rewrite-adj}.
    Note that adjoints are initialized before
    their first use.
\item[lines 9-11 and 12-13] similarly correspond to
    the differentiation of statements at lines $3$ and $2$
    in the original program.
\end{description}

Reverse-mode AD is exposed to the user by means of the
classical {\em vector-Jacobian product} (\fut{vjp})
interface:
\[
\textbf{vjp} : (f : \alpha \rightarrow \beta) \rightarrow (x : \alpha) \rightarrow (\overline{y} : \beta) \rightarrow \alpha 
\]

\fut{vjp} is a second-order function that computes the
derivative of $f$ at point $x$ given that the adjoint
of the result is $\adj{y}$. For example, the Jacobian
of $f$ at point $x$ can be computed by mapping 
\fut{vjp f x} on the unit vectors of the result type
$\beta$, i.e., each invocation of \fut{vjp} computes
one row of the Jacobian.
The notation in~\ref{eq:rewrite-adj} is a bit confusing:
since \fut{vjp} computes the vector-Jacobian product, it
should probably be written as 
$\adj{a} \ \ \mathrel{+}= \ \ \adj{v} \ \frac{\partial f(a,b,\ldots)}{\partial a}$,
as it makes a difference when \fut{v} is a vector.
We warn the reader that the rest of the paper will use
the established, albeit slightly confusing, notation. 

In the paper, we will denote by $\mathbb{VJP}$ the program
transformation (hinted in figure~\ref{fig:eg-revad-basic})
that implements reverse-mode AD and we
will use $\mathbb{VJP^{LAM}}$ to denote the code transformation
applied to the syntactic category of lambda functions.
(Other syntactic categories are, for example
expressions, statements and body of statements). 

\section{Preliminaries: Reverse-AD of Reduce}
\label{sec:reduce}

This section presents the algorithm (re-write rules) that
implements reverse-mode differentiation of \fut{reduce}: 
Section~\ref{subsec:red-gen} presents the ``general'' case,
in which the reduce operator is merely constrained to not
use any free variables,
Section~\ref{subsec:red-base} specializes the algorithm
to commonly-used operators addition, min/max, multiplication
and to vectorized operators, and Section~\ref{subsec:red-inv}
specializes the algorithm to a class of operators that are
commutative and (left-)invertible.

The goal of this section is to make the presentation self-contained,
since the rest of the paper builds on the type
of reasoning used for reduce. 
The content of this section is not a scientific contribution of
this paper because, for example, the general rule and specialized
cases were briefly presented elsewhere~\cite{futhark-ad-sc22} and
the specialization proposed in Section~\ref{subsec:red-inv} is not
implemented yet.
 
\subsection{General Algorithm}
\label{subsec:red-gen}

We start deriving the general rule from the definition of
\fut{reduce}. Given an associative operator $\odot$
with neutral element $e_\odot$, we have
\[
  y = \Reduce ~ \odot ~ e_\odot ~ [a_0, a_1, \ldots, a_{n-1}]
\]
which is equivalent to
\[
  y = a_0 \odot a_1 \odot \ldots \odot a_i \odot \ldots \odot a_{n-1}
\]
For each $a_i$, we can then group the terms of the reduce as:

\[
  y \ = \ \underbrace{a_0 ~ \odot \cdots \odot ~ a_{i-1}}_{l_i} ~ \odot ~ a_i ~ \odot ~ \underbrace{a_{i+1} ~ \odot \cdots \odot ~ a_{n-1}}_{r_i}
\]

Assuming $l_i$ and $r_i$ are known (i.e., already computed) and that
$\odot$ does not use any free variable then we can directly apply the
core rule for reverse AD, given in 
equation~\ref{eq:rewrite-adj}, to compute all the contributions
to the adjoints $\adj{a_i}$ in parallel:
\begin{equation} \label{eq:red-adj}
  \overline{a_i} ~ \mathrel{\overline{+}}= ~ \frac{\partial{(l_i \odot a_i \odot r_i)}}{\partial{a_i}} ~ \overline{y}
\end{equation}
\noindent where $\adj{y}$ --- the adjoint of the reduction result
$y$ --- has aleardy been determined by execution of the
return sweep until this point.

Here, $l_i$ and $r_i$ for all $i=0,\ldots,n-1$ can be computed
by two exclusive scans; one on the original array and one on the
reversed array. Exclusive scan is defined as:
\[
  \Scan^{exc} ~ \odot ~ e_\odot ~ [a_0, a_1, \ldots, a_{n-1}] \ \equiv \ [e_\odot,~a_0,~a_0\odot a_1, \ldots, a_0\odot\ldots a_{n-2}]
\]

\begin{figure}
\begin{lstlisting}[language=futhark, numbers=left]
-- Assuming array as of length n 
-- Primal trace is the same as original:
let y = reduce $\odot$ $e_\odot$ as

-- Return sweep:
let ls = $\Scan^{exc}$ $\odot$ $e_\odot$ as  -- forward exclusive scan
let rs = $\Reverse$ ad       -- reverse exclusive scan
       $\pipe$ $\Scan^{exc}$ $(\FnU{x ~ y}{y ~ \odot ~ x})$ $e_\odot$
       $\pipe$ $\Reverse$
    -- applying the core rule of reverse AD:
let $\adj{as}$ $\overline{\ttt{+}}\ttt{=}$ $\kw{map3}$ (\ $l_i$ $a_i$ $r_i$ -> $\adj{f_{l_i,r_i}}$ $a_i$) ls as rs 
    $\kw{denoting}$ $\adj{f_{l_i,r_i}} ~ \leftarrow ~ \mathbb{VJP^{LAM}} ~ (\FnU{ a_i }{l_i \odot a_i \odot r_i})~\adj{y}$
       
\end{lstlisting}\vspace{-2ex}
\caption{Reverse AD general re-write rule for \fut{reduce}}
   \label{fig:red-gen-rule}
\end{figure}

Figure~\ref{fig:red-gen-rule} shows the general-case
algorithm for reduction, where 
\begin{itemize}
\item $\pipe$ is the \emph{pipe operator} which composes
    functions left to right, 
\item $\adj{+}$ is a multidimensional addition
    (implemented as a tower of \fut{map}s)
    with rank equal to the rank of the input array, and
\item $\adj{f_{l_i,r_i}}$ is recursively obtained by applying
    the $\mathbb{VJP^{LAM}}$ transform to generate
    code corresponding to 
$\nicefrac{\partial{(l_i \odot a_i \odot r_i)}}{\partial{a_i}} \cdot \adj{y}$.
\end{itemize}

We observe that the transformation preserves the parallel
asymptotic of the original program --- since both \fut{reduce} and
\fut{scan} have linear work and logarithmic depth. However, it
incurs a rather large AD overhead since, even under aggressive
fusion, the differentiated code still requires about $8\times$
more global-memory accesses than the original reduce. %
Fortunately, standard operators, discussed next, admit more
efficient differentiation.

\subsection{Specialization for Common Operators}
\label{subsec:red-base}

Specialized rules for addition, min, max and multiplication
on numeric types are known~\cite{four-spec-cases} --- the rule
of multiplication in~\cite{four-spec-cases} uses two \fut{scan}s,
but a more efficient one is presented in~\cite{futhark-ad-sc22}.
This section recounts them in more detail for completeness. 

\subsubsection{Addition}
The primal trace of \lstinline{let y = reduce (+) 0 as} is
the same as the original and its return sweep adds $\adj{y}$
to each element of $\adj{as}$ since 
$\nicefrac{\partial{(l_i + a_i + r_i)}}{\partial{a_i}}$
simplifies to $1$, hence the return sweep is: 
\[
\Stm{\adj{as}}{\kw{replicate} ~n~ \adj{y} \ \ \pipe \ \ \kw{map2} \ (\ttt{+}) \ \adj{as}}
\]
We expect the AD overhead to be as high as $3\times$ since
the original code performs $n$ reads from global memory,
and the differentiated code performs $3\cdot n$ accesses: 
$2\cdot n$ reads and $n$ writes.\footnote{
  The {\tt reduce} of the primal performs $n$ reads, and the 
  {\tt map} of the return sweep performs $n$ reads and $n$ writes:
  The {\tt replicate} is fused with the {\tt map} and thus
  not considered, but the {\tt reduce} of the primal cannot
  be fused since its result {\tt y} is used in the {\tt map}.
}

\subsubsection{Min/Max} A reduction with min (max) selects
the minimum (maximum) element of an array. Assume that the
latter is located at position $k$. It follows that the
contribution to the adjoint of $a_i$ is:
\begin{itemize}
\item $0$ for any $i \neq k$ because the result
      \fut{y} does not depend on $a_i$,
\item $\adj{y}$ for the $k^{th}$ element, since
      $\frac{\partial min(a_0, \ldots, a_k, \ldots, a_{n-1}) }{\partial a_k} = \frac{\partial a_k}{\partial a_k} = 1$.
\end{itemize}

The primal trace is thus a lifted reduction whose 
associative and {\em commutative} operator, denoted
$\ttt{min}^{L}$, keeps track of the minimum value
together with its index --- in case of duplicates
we choose the smallest index corresponding to the
minimum value:
\[
\Stm{(k, y)}{\kw{zip} \ [0,\ldots,\ttt{n}-1] \ \ttt{as} \ \pipe \ \Reduce \ \ttt{min}^{L} \ (\ttt{n}, \ \infty)}
\]
The return sweep updates (only) the adjoint at position \fut{k}:
\[
\begin{array}{l}
\Stm{\adj{a_{min}}}{\kw{if} \ k < n \ \kw{then} \ \adj{as}[k] + \adj{y} \ \kw{else} \ 0_\alpha}\\
\Stm{\adj{as}}{\kw{scatter} \ \adj{as} \ \ \ttt{[~k~]} \ \ \ttt{[} \ \adj{a_{min}} \ \ttt{]}}
\end{array}
\]
The update is implemented in terms of the parallel-write operator
\fut{scatter}, which has the semantics that it discards the updates
of out of bounds indices (e.g., the case when \fut{as} is empty).

We expect the AD overhead to be around $2\times$: because the
original and primal both perform $n$ memory reads to compute
the reduction --- the index space $[0,\ldots,\ttt{n}-1]$ is fused, 
hence not manifested in memory --- and the return sweep might
need to initialize $\adj{as}$ (with zeroes), which requires
another $n$ memory writes. 
GPU implementations may suffer from (expensive) host-to-device
transfers if scalars and arrays are kept in the CPU and GPU memory
space, respectively, i.e., the values of \fut{k} and $\adj{as}[k]$
have to be brought from GPU to CPU, and the value of $\adj{a_{min}}$
has to be transferred back to GPU.

\begin{figure}
\begin{lstlisting}[language=futhark, numbers=left]
-- Original:
let y = reduce (*) 1 as
 
-- Primal trace:
let (n$^{=0}$, y$^{>0}$) =
  as $\pipe$ map (\a -> if a==0 then (1i64,1) else (0,a)) 
     $\pipe$ reduce (+, *) (0i64, 1)
let y = if n$^{=0}$ > 0 then 0 else y$^{>0}$

-- Return sweep:
let $\adj{as}$ = map2 (\ a $\adj{a}$ ->
                 $\adj{a}$ + if n$^{=0}$ == 0
                     then (y / a) * $\adj{y}$
                     else if n$^{=0}$ == 1 && a == 0
                          then y$^{>0}$ * $\adj{y}$
                          else 0
              ) as $\adj{as}$
\end{lstlisting}\vspace{-2ex}
\caption{Reverse AD rule for \fut{reduce} with multiplication}
   \label{fig:red-mult-rule}
\end{figure}

\subsubsection{Multiplication}
The quantity of interest is 
\[
\frac{\partial{y}}{\partial a_i} \ = \ \frac{\partial{(l_i \cdot a_i \cdot r_i)}}{\partial a_i} \ = \ l_i \cdot r_i
\]
If all elements would be known to be different than zero,
then $l_i \cdot r_i$ can be computed as $\nicefrac{y}{a_i}$,
resulting in the return sweep:
\[
  \Stmpb{\adj{as}}{\kw{map} \ (\lambda \ a_i \ \rightarrow \ (y / a_i) \ * \ \adj{y}) \ as}
\]
The case when some elements may be zero is treated
by extending the primal to compute (i) the number of
zero elements $n^{=0}$ and (ii) the product of the
non-zero elements $y^{>0}$.
Two additional cases require consideration:
\begin{itemize}
\item If exactly one element at index $i_0$ is zero, 
  then $l_i * r_i$ is zero for all other elements and
  only $\adj{a_{i_0}}$ is updated:
  $\overline{a_{i_0}} \mathrel{+}= y^{>0} * \overline{y}$.
\item If more than one zero exists, then $\overline{as}$
  remains unchanged.
\end{itemize}

Figure~\ref{fig:red-mult-rule} shows the code that implements
this algorithm. The reason for moving the $\kw{if}$ inside the
\fut{map} is to permit utilization of outer(-map) parallelism, 
in case it exists 
(otherwise the introduced control flow might prevent it). 
Assuming that in the common case \fut{as} does not contain
zeroes, the AD overhead can be as high as:
\begin{itemize} 
\item[$3\times$] if $\adj{as}$ is initialized at this
      point (i.e., lastly used in the original reduce),
      because its initialization will be fused
      with the \fut{map},
\item[$4\times$] otherwise, i.e., $2\cdot n$ reads and
      $n$ writes due the \fut{map} on the return sweep
      and $n$ reads due to the \fut{reduce} of the primal.
\end{itemize} 

\subsubsection{Vectorized Operators}
Vectorized operators are transformed with \textsc{Irwim}
re-write rule~\cite{futhark-fusion} that essentially
interchanges the outer reduce inside the inner map,
thus simplifying the reduce operator:
\begin{equation}\label{Irwim}
  \begin{array}{c}
  \Reduce \ (\kw{map2} \ \odot) \ (\kw{replicate} \ \ttt{n} \ e_\odot) \ \ttt{matrix} \\
     \equiv \\
  \Map \ (\Reduce \ \odot \ e_\odot) \ (\kw{transpose} \ \ttt{matrix})
  \end{array}
\end{equation}
The intuition is that summing up the elements on each column
of a matrix (\fut{reduce (map2 (+))}) is equivalent to
transposing the matrix, and summing up the elements on
each row (\fut{map (reduce (+))}). 
This is generalized to arrays of arbitrary rank by extending
$\kw{transpose}$ to permute the two outermost dimensions.

The \textsc{Irwim} rule is systematically applied 
in the $\mathbb{VJP}$ transformation whenever it matches
and differentiation is performed on the {\em resulting code}.
This is practically important because:
\begin{itemize}
\item it enables efficient differentiation of vectorized
      common operators without extra implementation effort;
\item some differentiation rules~\cite{PPAD} do not turn
      reductions with vectorized operators into 
      scans or reductions with vectorized operators --- such
      (arbitrary) operators on arrays are challenging to
      be mapped efficiently to the GPU hardware, e.g.,
      the code produced by Futhark 
      has abysmal performance.
\end{itemize}

\subsection{Specialization for Invertible Operators}
\label{subsec:red-inv}

The treatment of multiplication hints that the underlying
reasoning can be extended to encompass a larger class
of operators. The key properties that we have used are
\begin{description}
\item[commutativity:] in that $l_i \cdot a_i \cdot r_i$
    can be rewritten as $l_i \cdot r_i \cdot a_i$,
\item[invertibility:] in that knowing the current element
    $a_i$, the total number of zeros $n^{=0}$ and the
    product of non-zero elements $y^{>0}$, one can
    uniquely compute $l_i \cdot r_i$.
\end{description}
Similar to the work on near-homomorphisms~\cite{gorlatch:almost-hom},
we also observe that some non-invertible operations (such as
multiplication) can be inverted by extending them to
compute a (small) baggage of extra information.
This extra information makes the operator injective and
hence left invertible.

We propose to extend the language to allow the user
to connect an associative and commutative operator
$\odot : \alpha \rightarrow \alpha \rightarrow \alpha$ with 
\begin{itemize}
\item its lifted associative, commutative and (left-)invertible operator
      $\odot^L : \beta \rightarrow \beta \rightarrow \beta$ and its
      (left) inverse $\odot^L_{inv} : \beta \rightarrow \beta \rightarrow \beta$.\footnote{
  We argue that this extension is reasonable since parallel
  languages commonly \emph{assume} that the operators of reduce
  and scan are associative, and that those of reduce-by-index
  are also commutative. In fact, verification of such
  properties is undecidable in general.
}
\item a pair of functions that convert between $\alpha$ and $\beta$,
    denoted $f^{>}_{\odot} : \alpha \rightarrow \beta$, and
    $f^{<}_{\odot} : \beta \rightarrow \alpha$. 
\end{itemize}
The properties that the user must ensure to hold are:
\begin{itemize}
\item[(1)] $f^<_\odot \ \ \circ \ \ (\Reduce \ \odot^L \ e_{\odot^L}) \ \ \circ \ \ (\kw{map} \ f^>_\odot) \ \ \ \equiv \ \ \ \Reduce \ \odot \ e_\odot$
\item[(2)] for sanity, $e_{\odot^L}$ must belong to the co-domain of $f^>_\odot$,
\item[(3)] $\forall \ a, b$ if $z = a \ \odot^L \ b$
            then it holds that $a \ = \ z \ \odot^L_{inv} \ b$ and $b \ = \ z \ \odot^L_{inv} \ a$.
\end{itemize}
It can be derived from (1) and (2) that 
$e_{\odot^L} \equiv f^>_\odot(e_\odot$), that 
$e_\odot \equiv f^<_\odot(e_{\odot^L})$ and that
$f^<_\odot \circ f^>_\odot \ \equiv \ id$.

The lifted operators and conversions for multiplication are:
\[
\begin{array}{ccc}
(n_1^{=0}, \ y_1^{>0}) \ \cdot^{L} \ (n_2^{=0}, \ y_2^{>0}) & = & (n_1^{=0} + n_2^{=0}, \ y_1^{>0} \cdot y_2^{>0}) \\
(n_1^{=0}, \ y_1^{>0}) \ \cdot^{L}_{inv} \ (n_2^{=0}, \ y_2^{>0}) & = & (n_1^{=0} - n_2^{=0}, \ y_1^{>0} / y_2^{>0}) \\
\end{array}
\]
\[
\begin{array}{lcl}
f^{>}_{\cdot} \ (a_i) & = & \kw{if} \ (a_i == 0) \ \kw{then} \ (1, 1) \ \kw{else} \ (0, a_i) \\
f^{<}_{\cdot} \ (n^{=0}, y^{>0}) & = & \kw{if} \ (n^{=0} == 0) \ \kw{then} \ y^{>0} \ \kw{else} \ 0 
\end{array}
\]

Instead of differentiating $\Stm{y}{\Reduce \ \odot \ e_\odot \ as}$,
the $\mathbb{VJP}$ transform would be applied to the semantically
equivalent code:
\begin{equation}\label{eq:enabling-inv}
\begin{array}{l}
\Stm{as^L}{\kw{map} \ f^{>}_{\odot} \ as} \\
\Stm{y^{L}}{\Reduce \ \odot^L \ e_{\odot^L} \ as^L} \\
\Stm{y}{f^{<}_{\odot} \ y^L}
\end{array}
\end{equation}
and the code generation of the return sweep for the (middle)
reduce statement will exploit the commutativity and
invertibility of $\odot^L$:
\begin{lstlisting}[language=futhark, numbers=left]
let $\adj{as^L}$ $\adj{+}$=
  map2 (\ $a^L$ -> let $b^L$ = $y^L$ $\odot^L_{inv}$ $a^L$ in $\adj{\odot_{b^L}}^L$ $a^L$) $as^L$
    $\kw{denoting}$ $\adj{\odot_{b^L}}^L \ \leftarrow \ \mathbb{VJP^{LAM}} \ (\lambda \ x \ \rightarrow \ b^L \ \odot^L \ x) \ \adj{y^L}$ 
\end{lstlisting}
\noindent where $\adj{\odot_{b^L}}^L$ implements the
differentiation of $y^L \ = \ b^L \ \odot^L \ a^L$ with
respect to $a^L$ according to core re-write
rule~\ref{eq:rewrite-adj}, i.e.,
\[
\adj{\odot_{b^L}}^L \ (a^L) \ \ \equiv \ \ \frac{\partial (b^L \odot a^L)}{\partial a^L} \ \cdot \ \adj{y^L}
\]

\noindent An interesting example of associative, commutative
and invertible operator that we are unaware to have been
previously reported is:
\[
\begin{array}{ccc}
(p_1, \ s_1) \ \text{`sumOfProd`} \ (p_2, \ s_2) & = & (p_1 + p_2 + s_1 \cdot s_2, \ s_1 + s_2) \\
(p, s) \ \text{`sumOfProd`}_{inv} \ (p_2, \ s_2) & = & (p - p_2 - (s - s_2) \cdot s_2, \ s - s_2) \\
\end{array}
\]
For example, $\Sigma_{0\leq i<j<n} (a_i \cdot a_j)$ --- where
$a_i$ and $a_j$ denote different elements of an array $as$ of
length $n$ --- can be computed with:
\[
\begin{array}{l}
\kw{map} \ f^{>}_{\text{sumOfProd}} \ as \ \pipe \ \Reduce \ \text{sumOfProd} \ (0,0) \ \pipe f^{<}_{\text{sumOfProd}}
\end{array}
\]
and efficiently differentiated, as shown above. 
The conversions of the corresponding {\em near-homomorphism} are:
\[
\begin{array}{lcl}
f^{>}_{\text{sumOfProd}} \ (a_i) & = & (0, a_i) \\
f^{<}_{\text{sumOfProd}} \ (p, \ \_) & = & p 
\end{array}
\]

\section{Reverse-AD for Reduce-By-Index}
\label{sec:red-by-ind}

We recall that reduce-by-index, a.k.a., multi-reduce, is a
second-order array combinator that generalizes a histogram
computation~\cite{histo-sc20}: it reduces the values falling
in the same bin with an arbitrary associative and commutative
operator $\odot$, having neutral element $e_\odot$.
Its type and sequential/imperative semantics are: 
\begin{lstlisting}[language=futhark, mathescape=true]
def reduce_by_index $\forall$ w,n.
            (hist: *[w]$\alpha$) ($\odot : \alpha \rightarrow \alpha \rightarrow \alpha$) (e$_\odot$ : $\alpha$)
            (ks:  [n]int) (vs:  [n]$\alpha$) : *[w]$\alpha$ =
  for i = 0..n-1 do
      key = ks[i]
      if 0 <= key && key < w
        hist[key] = hist[key] $\odot$ vs[i]
  return hist
\end{lstlisting}

To simplify the reverse-mode AD transformation we systematically
re-write \fut{let} statements of the kind:
\begin{lstlisting}[language=futhark, numbers=none]
let hist = reduce_by_index hist$_0$ $\odot$ e$_\odot$ ks vs
\end{lstlisting}
into the semantically-equivalent code:
\begin{lstlisting}[language=futhark, numbers=none]
let xs = reduce_by_index (replicate w $e_\odot$)
                         $\odot$ e$_\odot$ ks vs
let hist = map2 $\odot$ hist$_0$ xs
\end{lstlisting}
that always applies \fut{reduce_by_index} to an initial
histogram consisting only of neutral elements $e_\odot$.
In this form, the adjoints of the initial histogram \fut{dst}
are decoupled from the \fut{reduce_by_index}, i.e., they are
updated by the differentiation of the \fut{map2}. This re-write
is reasonable because the design of \fut{reduce_by_index}
is built on the assumption that the histogram length is
(significantly) smaller than the length of the input.\footnote{
An asymptotic preserving re-write would be to replace the replicate
with a scatter that writes $e_\odot$ only at the positions
corresponding to the set of (unique) indices of $ks$.
Similarly, map can be replaced with a gather-scatter that
updates only those positions. 
}

\subsection{General Case}
\label{subsec:gen-case-red-by-ind}

We start by observing that reduce-by-index accepts a data-parallel
work-efficient $O(n)$ implementation obtained for example by (radix)
sorting the key-value pairs according to the keys (i.e., the indices
in $ks$), and
then by applying a segmented reduce to sum up (with $\odot$) 
each segment, where a segment corresponds to the (now
consecutive) elements that share the same key value.
%
%
A direct approach would be to apply the re-write above and
differentiate the resulted code. The rationale for not
taking this simple(r) path is because our differentiation of
{\tt scan} --- which appears in the implementation of segmented
reduce --- is not work preserving 
(see Section~\ref{subsec:scan-rule}). 

\begin{figure}
\begin{lstlisting}[language=futhark, numbers=left]
-- Assuming vs & ks of length n, and xs of length w
-- Primal trace is the same as original:
let xs = reduce_by_index (replicate w $e_\odot$)
                         $\odot$ $e_\odot$ ks vs
-- Return sweep:
let (sks,svs,siota) = zip3 ks vs [0..n-1] 
                    |> radixSortByFirst 
                    |> unzip3
let flag_fwd = [0 .. n-1]
             |> map (\i -> i==0 || sks[i-1]!=sks[i])
let flag_rev = [0 .. n-1]
             |> map (\i -> i==0 || flag[n-i])
let ls = seg_scan$^{exc}$ $\odot$ $e_\odot$ flag_fwd svs
let rs = reverse svs 
       |> seg_scan$^{exc}$ $\odot$ $e_\odot$ flag_rev
       |> reverse
let $\adjoint{svs}$ =
  map4 (\$k_i$ $v_i$ $l_i$ $r_i$ ->
         if $k_i$ < 0 || $k_i$ >= w
         then 0$_\alpha$ -- the zero of the element type $\alpha$
         else $\adj{f_{l_i,r_i}}$ $v_i$
           $\kw{denoting}$ $\adj{f_{l_i,r_i}} ~ \leftarrow ~ \mathbb{VJP^{LAM}} ~ (\FnU{ x }{l_i \odot x \odot r_i})~\adj{xs}[k_i]$
       ) sks svs ls rs
-- scratch creates an uninitialized array.
let $\adjoint{vs}$ = scatter (scratch $\alpha$ n) siota $\adjoint{svs}$
       |> map2 ($\adjoint{+}$) $\adjoint{vs}$
\end{lstlisting}\vspace{-2ex}
\caption{Reverse AD re-write rule for \fut{reduce_by_index}}
   \label{fig:histo-rule}
\end{figure}


Instead, the path we take builds on the one used in the general
case of reduce. Adapting equation~\ref{eq:red-adj} to reduce-by-index
results in:
\begin{equation} \label{eq:hist-adj}
\overline{v_i} \ \adjoint{+}= \ \frac{\partial \ (l_i \ \odot \ v_i \ \odot \ r_i)}{\partial \ v_i} \ \cdot \ \adjoint{x}_{k_i}
\end{equation}
where $v_i$ corresponds to \fut{vs[i]}, $l_i$ and $r_i$ correspond to
the forward and reverse partial sums (by $\odot$) of elements up to
position $i$ that have the same key $k_i$ as element $v_i$, and 
$\adjoint{x}_{k_i}$ is the adjoint of element at index $k_i$ in
the resulted histogram.
It follows that the implementation of ~\ref{eq:hist-adj} requires a
multi-scan for computing $l_i$ and $r_i$ within the segment corresponding
to elements sharing the same key $k_i$. This is commonly achieved by
(radix) sorting the key-value pairs according to the keys.
Figure~\ref{fig:histo-rule} shows the asymptotic-preserving re-write
rule that implements the reverse-mode differentiation:
\begin{description}
\item[lines 7-9] use a data-parallel implementation of radix sort
    ($O(n)$ work) to sort the key-value pairs (\fut{sks}, \fut{svs})
    according to the keys; the implementation only sorts $[0,\ldots,n-1]$
    according to keys, and then gathers $vs$ according to the
    resulted \fut{siota}.
\item[lines 10-11] create the flag array that semantically partitions
    the sorted values (\fut{svs}) into segments, such that all elements
    of a segment share the same key --- a \fut{true} value in the flag
    array correspond to the start of a segment. 
\item[line 14] performs a segmented scan, i.e., computes the (forward)
    prefix sum with operator $\odot$ for each segment.
\item[lines 12-13 and 15-17] compute in a similar way the reverse scan
    for each segment.
\item[lines 19-28] compute the adjoint contribution for each element
    of \fut{vs} (in sorted order): 
    if the index is out of the histogram bounds
        then the adjoint contribution is the zero value of the element
        type; otherwise 
    the $\mathbb{VJP}$ code transformation is applied
        to generate the code for differentiating the function
        $\lambda \ x \ \rightarrow \ l_i \ \odot \ x \ \odot \ r_i$
        at point $v_i$ given the adjoint of the result \fut{$\adj{xs}$[$k_i$]}
        (where $k_i$ denotes the key of $v_i$).
\item[lines 32-33:] the adjoint contributions are permuted to match
    the original ordering (\fut{scatter}) and added to the currently-known
    adjoint of \fut{vs} (which corresponds to uses of \fut{vs} after
    the \fut{reduce_by_index} operation).
\end{description}

The shown algorithm is asymptotic preserving --- all operations
have work $O(n)$ --- but it incurs a rather large (constant) overhead
due to sorting. Reasons are twofold: (1) a sorting-approach to implementing
generalized histograms has been shown to be several times slower
than Futhark's \fut{reduce_by_index}, even when using the state-of-the-art
implementation of CUB library, and (2) Futhark's radix sort implementation
we are using is between one-to-two order of magnitude slower than CUB's.

\subsection{Specialized Rules}
\label{subsec:histo-common}

\begin{figure}
\begin{lstlisting}[language=futhark, numbers=left]
let min$^L$ ($k_1$, $v_1$) ($k_2$, ($v_2$) =
  if      $v_1$ < $v_2$ then ($k_1$, $v_1$)
  else if $v_1$ > $v_2$ then ($k_2$, $v_2$)
  else (min $k_1$ $k_2$, v1)

-- Assuming vs & ks of length n, and xs of length w
-- Original for min operator:
let rep$_\infty$ = replicate w $\infty$
let xs = reduce_by_index rep$_\infty$ min $\infty$ ks vs
-- Primal trace for min operator:
let rep$_{(n,\infty)}$ = replicate w (n, $\infty$)
let (is$_{min}$,xs)= reduce_by_index rep$_{(n,\infty)}$ min$^L$ (n, $\infty$)
                      ks (zip [0,$\ldots$n-1] vs) $\pipe$ unzip
-- Return sweep for min operator:
let $\adj{vs}$ = map2 (\ i$_{min}$ $\adj{x}$ -> if $i_{min}$ >= n then 0$_\alpha$
                         else $\adj{vs}$[i$_{min}$] + $\adj{x}$
             ) is$_{min}$ $\adj{xs}$
      $\pipe$ scatter $\adj{vs}$ is$_{min}$
\end{lstlisting}\vspace{-2ex}
\caption{Reverse AD rule for \fut{reduce_by_index} with min.}
   \label{fig:hist-min-rule}
\end{figure}

The reasoning used for the specialized cases of reduce
also extends to reduce-by-index.


\subsubsection{Addition} 
Specializing general rule ~\ref{eq:hist-adj} to addition
results in:
\[
\overline{v_i} \ \ttt{+=} \ \frac{\partial \ (l_i \ + \ v_i \ + \ r_i)}{\partial \ v_i} \ \cdot \ \adjoint{x}_{k_i} \ \ \ \ \Rightarrow \ \ \ \ \overline{v_i} \ \ttt{+=} \ \adjoint{x}_{k_i}
\]
It follows that the primal remains identical with
the original:
\begin{lstlisting}[language=futhark, numbers=none]
let xs = reduce_by_index (replicate w 0) (+) 0 ks vs
\end{lstlisting}
and the return sweep adds to the adjoint of 
each element of \fut{vs} the adjoint of the histogram
element corresponding to its key:
\begin{lstlisting}[language=futhark, numbers=none]
let $\adj{vs}$ = map2 (\ k -> if k >= n then 0 else $\adj{xs}$[k]) ks
       $\pipe$ map2 (+) $\adj{vs}$
\end{lstlisting}
The AD overhead should be under a factor of $2\times$.
 
\subsubsection{Min/Max} Figure~\ref{fig:hist-min-rule}
shows the re-write rule. The primal trace still consists
of a reduce-by-index, but its operator is lifted to also
compute the index of the minimal element (\fut{min$^L$}).
The return sweep uses a scatter to update the adjoint of
only the element that has produced the minimal value for
that bin. In case of duplicates \fut{min$^L$} selects
the one at the smallest index in \fut{vs}.

\subsubsection{Multiplication}
Specializing rule ~\ref{eq:hist-adj} to multiplication yields:
\[
\overline{v_i} \ \ \ttt{+=} \ \ \frac{\partial \ (l_i \ \cdot \ v_i \ \cdot \ r_i)}{\partial \ v_i} \ \cdot \ \adjoint{x}_{k_i} \ \ \ \ \Rightarrow \ \ \ \ \overline{v_i} \ \ \ttt{+=} \ \ (l_i \ \cdot \ r_i) \cdot \ \adjoint{x}_{k_i}
\]
Computing $l_i \cdot r_i$ can be achieved in a similar way
as for reduction, by lifting the reduce by index (of the primal)
to compute for each bin the number of zero and the product of
non-zero elements falling in that bin. It follows that the lifted
operator is \fut{(i64.+, $\alpha$.*)} for some numeric type $\alpha$.
The return sweep consists of a map that adds the contributions
to the adjoint of \fut{vs}.

\begin{figure}
\begin{lstlisting}[language=futhark, numbers=left]
-- Assuming $\odot$ with inverse $\odot_{inv}$
-- Original is the same as primal:
let xs = reduce_by_index (replicate w $e_\odot$) $e_{\odot}$ ks vs

-- Return sweep:
let $\adj{vs}$ = map2(\$k_i$ $v_i$ -> let $b_i$= xs[k] $\odot_{inv}$ $v_i$ in $\adj{\odot_{b_i}}$ $v_i$)
            ks vs $\pipe$ map2 ($\adj{+}$) $\adj{vs}$
              $\kw{denoting}$ $\adj{\odot_{b_i}} \ \leftarrow \ \mathbb{VJP^{LAM}} \ (\lambda \ x \ \rightarrow \ b_i \odot x) \ \adj{xs}$[k]
              
-- A possible instantiation for $\odot$ and $\odot_{inv}$:
let sumOfProd (p1, s1) (p2, s2) =
      (p1 + p2 + s1*s2, s1 + s2)
let sumOfProd$_{inv}$ (p, s) (p2, s2) =
      (p - p2 - (s-s2)*s2, s - s2)
\end{lstlisting}\vspace{-3ex}
\caption{Reverse AD rule for reduce-by-index with left-invertible operator.}
   \label{fig-histo-inv}
\end{figure}

\subsubsection{Invertible Operators}
\label{subsubsec:inv-red-by-ind}
Similar to reduce, if the language would allow the user
to specify the (left-)inverse of an associative and
commutative operator, then a re-write similar to ~\ref{eq:enabling-inv}
would enable a significantly more efficient differentiation rule ---
illustrated in figure~\ref{fig-histo-inv} --- than the general case,
which is based on sorting. 

\subsubsection{Vectorized Operators}
Assuming $\odot : \alpha \rightarrow \alpha \rightarrow \alpha$,
$e_\odot : \alpha$, $\ttt{h}^0 : \ttt{[w][d]}\alpha$, $ks: [n]\kw{i64}$ and
$vss : \ttt{[n][d]}\alpha$ one can interchange the \fut{reduce_by_index}
inside the \fut{map} of a vectorized operator with the following rule:

\begin{equation}\label{Irbiwim}
  \begin{array}{l}
  \kw{reduce\_by\_index} \ \ttt{h}^0 \ (\kw{map2} \ \odot) \ (\kw{replicate} \ \ttt{d} \ e_\odot) \ \ttt{ks} \ \ttt{vss} \\
    \hspace{28ex}  \equiv \\
  \kw{map2} \ (\lambda~ \ttt{h}^0_{col} \ \ttt{vss}_{col} \rightarrow \\
  \hspace{10ex} \kw{reduce\_by\_index} \ \ttt{h}^0_{col} \ \odot \ e_\odot \ ks \ \ttt{vss}_{col}\\
  \hspace{6ex}) \ (\kw{transpose} \ \ttt{h}^0) \ (\kw{transpose} \ \ttt{vss}) \ \ \pipe \ \ \kw{transpose}
  \end{array}
\end{equation}

\noindent We currently do not utilize this rule prior for differentiation.
However, we do pattern match the case of vectorized common
operators --- i.e., a perfectly-nested sequence of maps which
ultimately applies plus, min/max or multiplication --- and
implement it as a special case: the primal will
correspond to a \fut{reduce_by_index} with the corresponding
vectorized lifted operators, and the return sweep would similarly
build a map nest to perform all necessary updates.

\subsubsection{Discussing Performance}\label{subsubsec:red-by-ind-discussion}
The design of the reduce-by-index construct~\cite{histo-sc20}
navigates the time-space tradeoff by employing
\begin{itemize}
\item a multi-histogram technique that is aimed at reducing
      the conflicts in shared (or global) memory by having
      groups of threads cooperatively building partial histograms, and 
\item a multi-pass technique that processes different partitions
   of the histogram at a time as a way to optimize trashing in
   the last-level cache.
\end{itemize}
In addition, reduce-by-index attempts to maintain the histogram(s)
in scratchpad memory if possible, and it implements the best
form of atomic update available on the hardware for the 
given datatype. For example (i) addition, min, max, multiplication
have efficient hardware implementation accessible through primitives 
such as \fut{atomicAdd}, (ii) datatypes that fit into 64-bits use
CAS instructions, while (iii) the rest use mutex-base locking,
which is quite expensive.

\enlargethispage{\baselineskip}

The AD overheads of the specialized cases of reduce-by-index 
(other than addition) are difficult to predict (or reason at
a high level) because of the lifted operators. For example,
\fut{int32.min} is efficiently supported in hardware, but
its lifting operates on \fut{(int64,int32)} tuples and thus
require mutex-based locking.  Furthermore, the size of the
element type is tripled, which restricts the multi-histogram
degree and thus impedes the reduction of conflicts. For vectorized
min, the tripling of the size might make it to not fit
in scratchpad memory anymore. All these factors may result
in a significant overhead that is not visible in the
re-write rules.

%
%
%


\section{Reverse-AD for Scan (Prefix Sum)}
\label{sec:scan}

Our differentiation of scan is restricted to
operators that are defined on tuples of scalars
of arbitrary dimension $d$, or to vectorized liftings
of such operators, i.e., a tower of maps applied
on top of such operators.
Our algorithm manifests and multiplies $d\times d$
Jacobians, which is arguably asymptotically preserving
since $d$ is a constant but it is not AD efficient.
Nevertheless, section~\ref{subsec:eval-scan} demonstrates
that it still offers competitive performance on 
many practical cases. 

Discussion is structured as follows:
section~\ref{subsec:deriving-scan} presents the
step-by-step rationale used to derive the algorithm --- which
we found interesting because it combines dependence
analysis on arrays with functional-style re-write
rules --- then section~\ref{subsec:scan-rule} puts
together the algorithm, and section~\ref{subsec:scan-spec}
presents several specializations that
enable significant performance gains --- e.g., 
vectorized operators and sparsity patterns --- and
concludes with a discussion that qualitatively
compare our algorithm with the one of PPAD~\cite{PPAD}.

\subsection{Deriving the Differentiation of Scan}
\label{subsec:deriving-scan}

An inclusive scan~\cite{segScan} computes all prefixes of
an array by means of an associative operator $\odot$ with
neutral element $e_\odot$:

\begin{lstlisting}[mathescape,basicstyle=\ttfamily\small] 
let rs = scan $\odot$ $e_\odot$[$a_0, \ldots, a_{n-1}$] 
       $\equiv$ [$a_0, a_0 \odot a_1,\ldots, a_0 \odot \ldots \odot a_{n-1}]$
\end{lstlisting}

While the derivation of (multi-) reduce builds on a functional-like
high-level reasoning, in scan's case, we found it easier to reason
in an imperative, low-level fashion. For simplicity we assume first
that $\odot$ operates on real numbers, and generalize later:

\begin{lstlisting}[language=C,mathescape,basicstyle=\small]
rs[0] = as[0]
for i in 1 $\ldots$ n-1 do
    rs[i] = rs[i-1] $\odot$ as[i]
\end{lstlisting}
%

The loop above that implements scan, writes each element of the
result array {\tt rs} exactly once. To generate its return sweep,
we can reason that we can fully unroll the loop, then
apply the main rewrite-rule from equation~\ref{eq:rewrite-adj} to
each statement and finally gather them back into the loop.
The unrolled loop is:
\begin{lstlisting}[mathescape,basicstyle=\small]
rs[0] = as[0]
rs[1] = rs[0] $\odot$ as[1]
$\ldots$
rs[n -1] = rs[n -2] $\odot$ as[n -1]
\end{lstlisting}
The application of $\mathbb{VJP}$ to each statement results in return sweep:
\begin{lstlisting}[mathescape,basicstyle=\small]
$\adjoint{rs}$[n-2] = $\nicediff{(\texttt{rs[n-2]} \odot \texttt{as[n-1]})}{\texttt{rs[n-2]}}$ * $\adjoint{rs}$[n-1]
$\adjoint{as}$[n-1] = $\nicediff{(\texttt{rs[n-2]} \odot \texttt{as[n-1]})}{\texttt{as[n-1]}}$ * $\adjoint{rs}$[n-1]
$\dots$
$\adjoint{rs}$[0] = $\nicediff{(\texttt{rs[0]} \odot \texttt{as[1]})}{\texttt{rs[0]}}$ * $\adjoint{rs}$[1]
$\adjoint{as}$[1] = $\nicediff{(\texttt{rs[0]} \odot \texttt{as[1]})}{\texttt{as[1]}}$ * $\adjoint{rs}$[1]
$\adjoint{as}$[0] = $\adjoint{rs}$[0]
\end{lstlisting}
The differentiated statements can be rolled back to form the loop:
\begin{lstlisting}[mathescape,basicstyle=\ttfamily\small]
$\adjoint{rs}$ = $\kw{copy}$ $\adjoint{ys}$
for i = n-1 $\ldots$ 1 do
    $\adjoint{rs}$[i-1] += $\nicediff{( \ttt{rs[i-1]} \ \odot \ \ttt{as[i]})}{\ttt{rs[i-1]}}$ * $\adjoint{rs}$[i]
    $\adjoint{as}$[i]   += $\nicediff{( \ttt{rs[i-1]} \ \odot \ \ttt{as[i]})}{\ttt{as[i]}}$ * $\adjoint{rs}$[i]
$\adjoint{as}$[0] += $\adjoint{rs}$[0]
\end{lstlisting}
where $\adjoint{ys}$ denotes the adjoint of \fut{rs} corresponding
to the uses of scan's result in the remaining of the program.
%
%


Simple dependence analysis, for example based on direction
vectors, shows that the loop can be safely distributed across
its two statements, since they are not in a dependency cycle:

\begin{lstlisting}[mathescape,basicstyle=\ttfamily\small]
$\adjoint{rs}$ = $\kw{copy}$ $\adjoint{ys}$
for i = n-1 $\ldots$ 1 do
    $\adjoint{rs}$[i-1] += $\nicediff{(\ttt{rs[i-1]} \ \odot \ \ttt{as[i]})}{\ttt{rs[i-1]}}$ * $\adjoint{rs}$[i]

for i = n-1 $\ldots$ 0 do
    $\adjoint{as}$[i] += (i==0) ? $\adjoint{rs}$[0] :
                     $\nicediff{(\ttt{rs[i-1]} \ \odot \ \ttt{as[i]})}{\ttt{as[i]}}$ * $\adjoint{rs}$[i]
\end{lstlisting}

The second loop exhibits no cross iteration dependencies,
hence the adjoints of \fut{as} can be computed by a \fut{map}.
%
The first loop can be expressed by the backwards linear
recurrence of form:
\begin{flalign*}
    &\overline{rs}_{n-1} = \overline{ys}_{n-1}
    \\
    &\overline{rs}_{i} = \overline{ys}_{i} + 
    cs_i \cdot \overline{rs}_{i+1}, i \in n-2\dots 0
\end{flalign*}
where $cs$ is defined by $cs_{n-1} = 1$ and $cs_{i}=\nicediff{(rs_i \odot as_{i+1})}{rs_i}$.
Such a recurrence is known to be solved with a scan whose operator
is linear-function composition~\cite{Blelloch1990PrefixSA}.

\subsection{Re-Write Rule for Arbitrary-Tuple Types}
\label{subsec:scan-rule}


\begin{figure}
\begin{lstlisting}[language=futhark, numbers=left]
-- We denote with n the length of as and with 
--    d the dimensionality of the element type $\alpha$
-- Primal trace is the same as the original:
let rs = scan $\odot$ $e_{\odot}$ as

-- Return sweep:
--   (1) computes cs (Jacobians):
let cs = map (\i -> if i == n-1 
                   then $\mathbf{I_d}$  
                   else $\ola{J}_{\odot,\ttt{as}}$ i rs[i]
                -- ^ i.e., $\nicediff{(\ttt{rs[i]} ~\odot~ \ttt{as[i+1]})}{\ttt{rs[i]}}$
             ) [0,$\ldots$,n-1]
  $\kw{denoting}$ $\ola{J}_{\odot,\ttt{as}} \ i \ x \ \ = \ \ (f_{0,i,\odot,\ttt{as}}~x, \ \ldots,~f_{d-1,i,\odot,\ttt{as}}~x)$
        $f_{k,i,\odot,\ttt{as}} \ \leftarrow \ \mathbb{VJP^{LAM}} \ (\lambda x \rightarrow x \odot \ttt{as[i+1]}) \ \ttt{(unitVec k)}$

--   (2) computes the adjoint of rs by means of
--       parallelizing a backward linear recurrence
let lin$_o$ (b1,c1) (b2,c2) = (b2 + c2 $\cdot$ b1, c2 $\times$ c1)
let ($\adjoint{rs}$, _) = zip (reverse $\adjoint{ys}$) (reverse cs) 
            $\pipe$ scan lin$_o$ ($\ttt{0}_d, \ \mathbf{I_d}$) 
            $\pipe$ reverse $\pipe$ unzip
--   (3) updates the adjoint of as by a map:
let $\adjoint{as}$ $\adjoint{+}$=
  map (\i $\adjoint{ri}$ ai ->
         if i == 0 then $\adjoint{ri}$
         else $g_{i,\odot}$ ai -- i.e., $\nicediff{(\texttt{rs[i-1]} \odot \texttt{ai})}{\texttt{ai}} ~\cdot~ \adjoint{ri}$
         $\kw{denoting}$ $g_{i,\odot} \ \leftarrow \ \mathbb{VJP^{LAM}} \ (\lambda x \rightarrow \ttt{rs[i-1]} ~\odot ~x) \ \adjoint{ri}$
      ) [0,$\ldots$,n-1] $\adjoint{rs}$ as

\end{lstlisting}\vspace{-2ex}
\caption{Reverse-AD Rule for Scan.}
   \label{fig-scan-gen}
\end{figure}

The reasoning used in the previous section generalizes
to $d$-dimensional tuples (chosen for simplicity) of
the same numeric type $\alpha$,
essentially by lifting scalar addition and multiplication
to operate on vectors and matrices.\footnote{
The reasoning generalizes also to tuples of heterogeneous 
scalar types.
}
Figure~\ref{fig-scan-gen} presents the proposed re-write
rule. 

For example, the linear-function composition operator has type
$\ttt{lin}_o : (\alpha^d, \alpha^{d\times d}) \rightarrow (\alpha^d, \alpha^{d\times d}) \rightarrow (\alpha^d, \alpha^{d\times d})$
and \fut{+}, $\cdot$ and $\times$ denote vector addition,
vector-matrix and matrix-matrix multiplication, respectively,
where vectors live in $\alpha^d$ (its zero is {\tt 0$_d$})
and matrices in $\alpha^{d\times d}$.
Similarly, $cs_{i>0}$ are the $d\times d$ Jacobians corresponding
to $\nicediff{(rs_i \odot as_{i+1})}{rs_i}$ and $cs_0 = \mathbf{I_d}$
is the identity matrix. 

The code for computing $cs_{i>0}$ --- represented
in figure~\ref{fig-scan-gen} by means of
$\ola{J}_{\odot,\ttt{as}}~\ttt{i rs[i]}$ --- is generated by
applying the $\mathbb{JVP^{LAM}}$ transformation to lambda
$\lambda x \rightarrow x \odot as[i+1]$ and to each of
the unit vectors ({\tt unitVec $k,~k=0\ldots, d-1$}) as
the adjoint of the result. 

The generated code consists of two kernels: one corresponding
to the fusion of the map computing \fut{cs} together with
the reversion of \fut{cs} and $\adj{ys}$ and the scan, and
the second corresponding to the fusion of the reversion of
$\adj{rs}$ with the map that updates $\adj{as}$.

\subsection{Specializations}
\label{subsec:scan-spec}

\subsubsection{Addition} It is folklore knowledge that
the return sweep of
%
\begin{center} $\Stm{ys}{\kw{scan} ~\ttt(+)~\ttt{0}~\ttt{as}}$ \hspace{3ex} is: \end{center}
\[
\Stm{\adj{as}}{\kw{scan}~\ttt{(+)}~\ttt{0}~\ttt{(}\kw{reverse}~\adj{ys}\ttt{)} \ \ \pipe \ \ \kw{reverse} \ \ \pipe \ \ \kw{map2}~\ttt{(+) 0} \ \adj{as}}
\]
This can also be derived from figure~\ref{fig-scan-gen}: 
$\nicediff{(\ttt{rs[i]} ~+~ \ttt{as[i+1]})}{\ttt{rs[i]}}$ simplifies
to $1$, hence \fut{cs = replicate n 1}, which means that we are
composing linear functions of the form 
$f_{\adj{ys}_i} ~x = \adj{ys}_i + x$, which results
in  
$~~\adj{rs} \ = \ \kw{scan}~\ttt{(+)}~\ttt{0}~\ttt{(}\kw{reverse}~\adj{ys}\ttt{)} \ \ \pipe \ \ \kw{reverse}$,
and so on.

\subsubsection{Vectorized Operators}
Scans with vectorized operators are transformed to
scans with scalar operators (whenever possible)
by the (recursive) application of the \textsc{Iswim} 
rule\footnote{
\textsc{Iswim} states that summing up the
elements of each column of a matrix can be achieved by
transposing the matrix, summing up each row and transposing
back the result. 
}~\cite{futhark-fusion}: 
\begin{equation}\label{Iswim}
  \begin{array}{c}
  \Scan \ (\kw{map2} \ \odot) \ (\kw{replicate} \ \ttt{n} \ e_\odot) \ \ttt{matrix} \\
     \equiv \\
  \Map \ (\Scan \ \odot \ e_\odot) \ (\kw{transpose} \ \ttt{matrix}) \ \ \pipe \ \ \kw{transpose}
  \end{array}
\end{equation}
and differentiation is applied on the resulted code.
This is essential because the ``general-case'' rule
in figure~\ref{fig-scan-gen} is {\em not} asymptotic
preserving in the case of array datatypes due to the
explicit manifestation and multiplication of Jacobians.

\subsubsection{Block-Diagonal Sparsity (BDS)}
The expensive step in our re-write rule of figure~\ref{fig-scan-gen}
is that entire $d\times d$ Jacobians corresponding to 
$\nicediff{(\ttt{rs[i]} ~+~ \ttt{as[i+1]})}{\ttt{rs[i]}}$ 
(computed at line $10$) are stored in \fut{cs} 
and later multiplied inside (the scan with operator)
$\ttt{lin}_o$ (line $18/20$).
It is not only that $\ttt{lin}_o$ takes $O(d^3)$ time, 
but more importantly, the size of the elements being
scanned is proportional with  $d^2$, which quickly
restricts (i) the amount of efficient sequentialization,
and ultimately (ii) the storing of intermediate data in
scratchpad memory that is paramount for the GPU efficiency
of scan. 

In this sense, we have implemented (compiler) analysis
to statically detect sparse Jacobians of block-diagonal
form. More precisely, we consider the case of $k$ blocks,
where each block has size $q\times q$, hence $d = k\cdot q$.
Multiplication preserves the block-diagonal shape: 
\[
\begin{bmatrix}
  \mathbf{M^1_1} & \cdots & 0 \\
  \vdots & \ddots & \vdots \\
  0 & \cdots & \mathbf{M^1_k}
\end{bmatrix}
\mathbf{\times}
\begin{bmatrix}
  \mathbf{M^2_1} & \cdots & 0 \\
  \vdots & \ddots & \vdots \\
  0 & \cdots & \mathbf{M^2_k}
\end{bmatrix} 
\mathbf{=} 
\begin{bmatrix}
  \mathbf{M^1_1}\times\mathbf{M^2_1} & \cdots & 0 \\
  \vdots & \ddots & \vdots \\
  0 & \cdots & \mathbf{M^1_k}\times\mathbf{M^2_k}
\end{bmatrix} 
\]

We multiply such a matrix with $v$, a vector of
length $d$ as such:
\vspace{-1ex}
\[
\begin{bmatrix}
\mathbf{v_1}, & \cdots, & \mathbf{v_k}
\end{bmatrix}
\mathbf{\times}
\begin{bmatrix}
  \mathbf{M_1} & \cdots & 0 \\
  \vdots & \ddots & \vdots \\
  0 & \cdots & \mathbf{M_k}
\end{bmatrix}
\mathbf{=} 
\begin{bmatrix}
\mathbf{v_1} \times \mathbf{M_1}\\
\vdots \\
\mathbf{v_k} \times \mathbf{M_k}
\end{bmatrix}
\]

If \fut{cs} has the BDS pattern, then, semantically, we
shrink its representation down to a tuple of $k$ arrays
each of dimension $n\times(q\times q)$ and similarly, 
$\adj{ys}$ to a tuple of $k$ arrays of dimension 
$n\times q$.\footnote{
  Since the original scan operator is defined
  on tuples of scalars and Futhark compiler uses
  a tuple of array representation, it follows that
  in practice, {\tt cs} is represented as 
  $k\cdot q\cdot q$ arrays
  of length $n$, that we tuple differently at no 
  runtime overhead. Similar thoughts apply to
  $\adj{ys}$ and $\ttt{lin}^{BDS}_o$, i.e., $\ttt{lin}^{BDS}_o$
  still operates on tuples of scalars.
}
The computation of $\adjoint{rs}$ is performed with
$k$ different scans, each of them using a scaled-down
(adjusted) operator $\ttt{lin}_o^{BDS}$ that is
semantically defined on elements of type 
$(\alpha^q, \alpha^{q\times q})$ --- the corresponding
vector- and matrix-matrix multiplications inside 
$\ttt{lin}_o^{BDS}$ are performed as shown above.
This reduces the element size of the scanned
array, enabling better utilization of scratchpad memory.

\subsubsection{Redundant Block-Diagonal (RBDS) Sparsity}\label{rbd-sparsity}
The case when the block-diagonal sparsity has the
additional property that all the blocks hold identical
values, i.e., $M_{1} = M_{2} = \ldots = M_{k}$,
allows an even more efficient implementation:
The representation
of \fut{cs} is shrunk down to (semantically) one
array of dimension $n\times(q\times q)$ and only
one scan is performed.
$\ttt{lin}_o^{RBDS}$ now operates on elements of type 
$(\alpha^d, \alpha^{q\times q})$ and it performs
\textbf{\em one multiplication of $q\times q$ matrices}, and
$k$ vector-matrix multiplications $\mathbf{V^q}\times M^{q\times q}$:
 
\[
\begin{bmatrix}
\mathbf{v_1}, & \cdots, & \mathbf{v_k}
\end{bmatrix}
\mathbf{\times}
\begin{bmatrix}
  \mathbf{M} & \cdots & 0 \\
  \vdots & \ddots & \vdots \\
  0 & \cdots & \mathbf{M}
\end{bmatrix}
\mathbf{=} 
\begin{bmatrix}
\mathbf{v_1} \times \mathbf{M}\\
\vdots \\
\mathbf{v_k} \times \mathbf{M}
\end{bmatrix}
\]

RBD sparsity has important applications:
differentiating the multiplication of two
$q\times q$ matrices $A \times B$ with
respect to $A$ (or $B$), results in a
Jacobian that consists of $q$ blocks of
size $q\times q$, in which each block is
equal to $B$ (or $A$), i.e., the Jacobian of
$\nicediff{(A \times B)}{A}$ is:
\[
\begin{bmatrix}
  B & \cdots & 0 \\
  \vdots & \ddots & \vdots \\
  0 & \cdots & B
\end{bmatrix}
\] 

This is important because the scan with $q \times q$
matrix multiplication is commonly used to parallelize
linear recurrences of degree $q$~\cite{Blelloch1990PrefixSA}, 
i.e., $x_i = a^0_i + a^1_i*x_{i-1} + \ldots + a^q_i*x_{i-q+1}$.

Similarly, differentiating the (classical) linear function
composition with respect to the first argument
also results in RBS sparsity, i.e., 
$\nicediff{(b_2+c_2\cdot b_1, c2 \cdot c1)}{(b_1,c_1)}$
has Jacobian
$
\begin{bmatrix}
  c_2 & 0 \\
  0 & c_2
\end{bmatrix}
$

\subsubsection{Discussion}
Our ``general-case'' rule for scan computes and multiplies
$d\times d$ Jacobians.
While this arguably preserves the work asymptotic
($d$ is a constant for tuples), it is not AD efficient\footnote{
There is no constant factor independent of the program that
bounds the AD overhead. 
} and it is theoretically inferior to the re-write
rule of PPAD~\cite{PPAD} --- shown in figure~\ref{fig:ppad-scan}
in Appendix --- which uses only the $\mathbb{VJP}$ transformation, 
thus avoiding operating with (full) Jacobians.

Our ``general-case'' rules is however faster than PPAD when
the dimensionality $d$ is one or two, and the vectorized-operator
and RBDS specializations makes it also more effective 
on many operators of practical interest, with
speed-ups commonly ranging from $1.3\times - 2.25\times$,
as reported in section~\ref{subsec:eval-scan}.
In particular the PPAD rule transforms a vectorized operator into
an un-vectorized one, which is challenging to map efficiently to
the GPU hardware --- especially when the dimensions of the array-based
element type are not statically known.  In Futhark's case, such
non-vectorized operators incur prohibitive AD overheads, e.g.,
two orders of magnitude.

Finally, for the un-vectorized operators that use array arguments,
we postulate that a more suited strategy would be to differentiate
the classical work-preserving (two-stage) implementation of
scan~\cite{segScan} written in terms of loop, map and scatter
operators. We plan to use PPAD's rule for high-dimensional
tuples of scalars (of arity larger than $3$) that do not
fall under the RBDS pattern.


\section{Experimental Evaluation}

The discussion is structured as follows:
section~\ref{subsec:eval-op-data-method} presents
the evaluation methodology and 
sections~\ref{subsec:eval-reduce}~,\ref{subsec:eval-scan}~and~\ref{subsec:eval-red-by-ind}
evaluate the performance of reverse-mode differentiation
of reduce, scan and reduce-by-index, respectively.

\subsection{Operators, Datasets, Methodology}
\label{subsec:eval-op-data-method}

\begin{figure}
\begin{lstlisting}[language=futhark, numbers=left]
def linFnComp (b1:f32, c1:f32) (b2:f32, c2:f32) =
      (b2 + c2*b1,    c2 * c1)
def sumOfProd (p1:f32, s1:f32) (p2:f32, s2:f32) =
      (p1 + p2 + s1*s2,    s1 + s2)
      
def matMul2x2 (a1:f32, b1:f32, c1:f32, d1:f32)
              (a2:f32, b2:f32, c2:f32, d2:f32) =
      ( a1*a2 + b1*c2,    a1*b2 + b1*d2
      , c1*a2 + d1*c2,    c1*b2 + d1*d2 )
-- ^ matMul3x3, and matMul5x5 are similarly defined

def satAdd (x: f32) (y: f32) : f32 =
      if (x+y) > 1000000 then 1000000 else x+y
\end{lstlisting}\vspace{-2ex}
\caption{Non standard operators used in evaluation.}
   \label{fig-eval-ops}
\end{figure}

The evaluation uses randomly generated arrays and 
single-precision float as the base numeric type.
\footnote{
  Due to negligence, we have used $32$-bit integer in some
  cases: we will fix this in a final version, but this should
  not significantly influence the performance.
} 

\subsubsection{Operators} The evaluated operators are: 
(i)   standard addition ($+$), multiplication (*), and min, 
(ii)  their vectorized forms, e.g., \fut{map2 (*)},
(iii) $2\times 2$, $3\times 3$ and $5\times 5$ matrix
      multiplication, e.g., which is used in the parallel
      implementation of linear recurrences,
(iv)  linear function composition,
(v)   sum of products, and
(vi)  saturated addition.
For convenience, figure~\ref{fig-eval-ops} shows the non-standard
ones. Of note, linear function composition and matrix multiplication
are only associative but not commutative, hence they are not valid
operators for reduce-by-index, which requires commutativity.

\subsubsection{Datasets} 
The evaluation of reduce and scan uses two datasets denoted
by $D_1$ and $D_2$:
\begin{description}
\item[\textbf{map2 (*)}:] 
  For vectorized multiplication, $D_1$ and $D_2$
  correspond to arrays of dimensions 
  $10^6 \ \times \ 16 $ and $10^7 \ \times \ 16$, respectively,
  which are provided in transposed form.
\item[\textbf{matMul5x5}:] For $5\times 5$ matrix multiplication,
  $D_1$ corresponds to $10$ million elements -- each element
  is a tuple of arity $25$ -- and $D_2$ corresponds to
  $50$ million elements.
\item[For the others] $D_1$ and $D_2$ correspond to $10$ and
  $100$ million elements, e.g., the element for sum-of-products
  is a tuple of floats (arity $2$).
\end{description}

The evaluation of reduce-by-index uses six datasets 
denoted by $D_{i,j}$ where $i=1,2$ refers to the length
of the input arrays and  $j=1,2,3$ refers to the length 
of the histogram: $31$, $401$ and $50000$ elements,
respectively. The length of the input arrays are
the same as before, except for the case of vectorized
operators where $D_{1,j}$ and $D_{2,j}$ correspond to
arrays of dimensions 
$10^6 \ \times \ 10 $ and $10^7 \ \times \ 10$, respectively.

\subsubsection{Hardware} The evaluation uses an Nvidia
A100 40GB PCIe GPU, which has the listed peak memory
bandwidth of $\mathbf{1555}$ \textbf{Gb/sec}.

\subsubsection{Methodology}
We measure the total application running time, but
{\em excluding} the time needed to transfer the
program input and (final) result between device
and host memory spaces. We report the average of
at least 25 runs --- or as many as are needed for
a $95\%$ confidence interval to be reached.

The performance of the primal (original program)
is reported as memory throughput, measured in
 \textbf{Gb/sec}. Denoting with $n$ the length
 of the input array and with $\beta$ the size of
 the array element type, the total number of bytes
 {\tt Nbytes} is computed as follows:
\begin{equation}\label{num-bytes}
\ttt{Nbytes} = 
\begin{cases}
            n \cdot \ttt{sizeof}(\beta), & \text{for reduce}\\
    2 \cdot n \cdot \ttt{sizeof}(\beta), & \text{for scan} \\
    3 \cdot n \cdot \ttt{sizeof}(\beta) \ + \ n \cdot 8, & \text{for reduce-by-index}
\end{cases}
\end{equation}
For reduce and scan these are the minimal number of bytes
that needs to be accessed from global memory, e.g., reduce
needs to read each element once. For reduce by index we
reason that:
\begin{itemize}
\item[(1)] reading the input array requires 
      $n \cdot \ttt{sizeof}(\beta)$ bytes,
\item[(2)] reading the key requires $n\cdot 8$ bytes,
      because the key is represented as a $64$-bit
      integer,
\item[(3)] updating the histogram {\em may} require
      a read and a write access, hence another
      $2 \cdot n \cdot \ttt{sizeof}(\beta)$ bytes.
\end{itemize}

For the primal, histograms of sizes $31$ and $401$
typically fit in scratchpad (shared/fast) memory, 
but histograms of size $50000$ do not and are stored
in global memory.
It follows that we choose to consider the
accesses that update the histogram 
in order to be able to meaningfully 
compare across different datasets and
implementations --- i.e., our measure of
memory throughput is essentially a normalized
runtime.
The consequence is that on small histograms
the reported \textbf{Gb/sec} may exceed the peak
memory bandwidth of the hardware, because the
histogram is maintained in shared memory.

The performance of the differentiated code ---
that computes both the primal and the adjoint
results --- is presented in terms of AD overhead,
which is defined as the ratio between the running
times of the derivative and primal 
(original) -- the lower the better.

\begin{figure*}
\hspace{-7ex}
\begin{minipage}[b]{.53\textwidth}
  \small\centering
  \begin{tabular}{l|rrr|rrr|rrr|}
   \textbf{Op} & \multicolumn{3}{c|}{\fut{reduce (+)}} & \multicolumn{3}{c|}{\fut{reduce min}} & \multicolumn{3}{c|}{\fut{reduce (*)}} \\
               & \textbf{Prim} & \textbf{Our} & \textbf{Cmp} & \textbf{Prim} & \textbf{Our} & \textbf{Cmp}   & \textbf{Prim} & \textbf{Our} & \textbf{Cmp} \\\hline
               & \textbf{Gb/s} & $\AdOv{AD-F}$ & $\AdOv{AD-C}$ & \textbf{Gb/s} & $\AdOv{AD-F}$ & $\AdOv{AD-C}$ & \textbf{Gb/s} & $\AdOv{AD-F}$ & $\AdOv{AD-C}$ \\\hline
    $\mathbf{D_1}$& $755$         & $2.4\times$&             & $755$         & $2.8\times$&         &  $755$      & $2.7\times$        & $9.1\times$        \\
    $\mathbf{D_2}$& $1246$        & $2.5\times$&             & $1261$        & $2.6\times$&         &  $1270$     & $2.9\times$        & $13.1\times$       \\\hline\hline
  \end{tabular}
  \vfill
  \bigskip
  \small\centering
  \begin{tabular}{l|rrr|rrr|rrr|}
   \textbf{Op} & \multicolumn{3}{c|}{\fut{reduce linFnComp}} & \multicolumn{3}{c|}{\fut{reduce sumOfProd}} & \multicolumn{3}{c|}{\fut{reduce (map2(*))}} \\\hline
    $\mathbf{D_1}$& $530$         & \hphantom{$0$}$5.0\times$&   $7.3\times$          & $808$         & $7.2\times$& $10.9\times$        &  $1016$       & \hphantom{$0$}$3.4\times$        & $668\times$    \\
    $\mathbf{D_2}$& \hphantom{$0$}$712$        & $5.9\times$&    $9.3\times$         & $1317$        & \hphantom{$0$}$10.\times$& $16.8\times$        &  $1342$       & $4.0\times$        & $797\times$   \\\hline\hline
  \end{tabular}
  \vfill
  \bigskip  
  \small\centering
  \begin{tabular}{l|rrr|rrr|rrr|}
   \textbf{Op} & \multicolumn{3}{c|}{\fut{reduce matMul2x2}} & \multicolumn{3}{c|}{\fut{reduce matMul3x3}} & \multicolumn{3}{c|}{\fut{reduce matMul5x5}} \\\hline
    $\mathbf{D_1}$& \hphantom{$0$}$611$  & \hphantom{$0$}$4.6\times$&  $9.5\times$  & \hphantom{$0$}$684$ & \hphantom{$0$}$6.4\times$& $22.3\times$ & \hphantom{$0$}$252$  & \hphantom{$0$}$6.3\times$ & $33.3\times$        \\
    $\mathbf{D_2}$& $844$  & $6.2\times$&  $12.4\times$ & $839$ & $7.6\times$& $27.0\times$ & $268$  & $6.7\times$ & $35.3\times$       \\\hline\hline
  \end{tabular}
  \vfill
\end{minipage}
\begin{minipage}[b]{.44\textwidth}
  \small\centering
  \begin{tabular}{rrr|rrr|rrr|}
  \multicolumn{3}{c|}{\fut{scan (+)}} & \multicolumn{3}{c|}{\fut{scan min}} & \multicolumn{3}{c|}{\fut{scan (*)}} \\
  \textbf{Prim} &\textbf{Our} &\textbf{Cmp} &\textbf{Prim} &\textbf{Our} &\textbf{Cmp} &\textbf{Prim} &\textbf{Our} &\textbf{Cmp}\\\hline
  \textbf{Gb/s} & $\AdOv{AD-F}$ & $\AdOv{AD-C}$ & \textbf{Gb/s} & $\AdOv{AD-F}$ & $\AdOv{AD-C}$ & \textbf{Gb/s} & $\AdOv{AD-F}$ & $\AdOv{AD-C}$ \\\hline
  $584$  & $1.8\times$& \hphantom{$3.3\times$} & $808$  & $2.5\times$& $5.2\times$ &  $808$  & $3.4\times$ & $4.8\times$ \\
  $1131$ & $2.8\times$& \hphantom{$5.7\times$} & $1111$ & $2.8\times$& $6.3\times$ &  $1131$ & $4.1\times$ & $6.0\times$ \\\hline\hline
  \end{tabular}
  \vfill
  \vfill
\bigskip
  \small\centering
  \begin{tabular}{rrr|rrr|rrr|}
    \multicolumn{3}{c|}{\hphantom{\textbf{00}}\fut{scan linFnComp}} & \multicolumn{3}{c|}{\hphantom{\textbf{00}}\fut{scan sumOfProd}} & \multicolumn{3}{c|}{\hphantom{\textbf{00}}\fut{scan (map2(*))}} \\\hline
  $982$  & \hphantom{$0$}$3.4\times$& $6.2\times$ & $970$  & \hphantom{$0$}$4.5\times$ & $5.9\times$ &  $105$ & \hphantom{$0$}$0.44\times$ & $33.4\times$ \\
  $1208$ & $3.8\times$& $7.0\times$ & $1202$ & $5.3\times$ & $6.8\times$ &  $114$ & $0.43\times$ & $32.8\times$ \\\hline\hline
  \end{tabular}
  \vfill
\bigskip
  \small\centering
  \begin{tabular}{rrr|rrr|rrr|}
    \multicolumn{3}{c|}{\hphantom{\textbf{00}}\fut{scan matMul2x2}} & \multicolumn{3}{c|}{\hphantom{\textbf{00}}\fut{scan matmul3x3}} & \multicolumn{3}{c|}{\hphantom{\textbf{00}}\fut{scan matMul5x5}} \\\hline
  $1032$ & \hphantom{$0$}$4.3\times$& $6.9\times$ & $880$  & \hphantom{$0$}$7.6\times$ & $14.8\times$ &  $192$ & \hphantom{$0$}$6.9\times$ & $12.9\times$ \\
  $1149$ & $4.6\times$& $7.3\times$ & $943$  & $8.1\times$ & $15.7\times$ &  $191$ & $6.9\times$ & $12.8\times$ \\\hline\hline
  \end{tabular}
  \vfill
\end{minipage}\vspace{-1ex}
\caption{Performance of the reverse-mode differentiation of 
         \fut{reduce} and \fut{scan}.
         The performance of the original program (primal) is measured
         in \textbf{Gb/sec} and is reported in column \textbf{Prim}.
         AD performance is reported in terms of AD Overhead, which is
         defined as the ratio between the differentiated  and primal
         runtimes (the lower the better).  The AD overhead of our 
         approach is reported in column \textbf{Our} and of the
         competitor technique in column \textbf{Cmp}.
         The competitor are the re-write rules of PPAD~\cite{PPAD} except
         from reduce with multiplication, where we use the general
         case of differentiating reduction, which is similar to ~\cite{four-spec-cases}.
}
\label{fig:eval-red-scan-perf}
\end{figure*}

\subsection{Reduce}
\label{subsec:eval-reduce}

The left-hand side of figure~\ref{fig:eval-red-scan-perf} presents the
performance of the primal and reverse-mode differentiation of reduce
on the evaluated operators and datasets. As competitor, we use the
algorithm presented in PPAD~\cite{PPAD}, except for multiplication,
where we use our general-case as competitor, which is also somewhat
similar to ~\cite{four-spec-cases} and is more efficient than PPAD.
We make the following observations:
\begin{itemize}
\item In most cases, e.g., $+, \cdot, \ttt{min}$, $D_1$
      is too small to overcome the ``system'' overheads, resulting
      in sub-optimal performance of about half the peak bandwidth,
      e.g., the total runtime is about $53$ micro-seconds $\mu{}s$
      for min, and launching the kernel takes ten(s) $\mu{s}$.
      Matter are much improved on the larger $D_2$. 

\item The specializations for $+, \cdot, \ttt{min}$ enable 
      efficient differentiation --- all AD overheads are
      under $3\times$. The case of multiplication highlights
      the impact of specialization: it offers $3.4-4.5\times$
      speedup in comparison with the general case (\textbf{Cmp}).

\item linFnComp and sumOfProd are treated with the general-case
      algorithm and result in significantly larger overheads
      $5-10\times$.
      
\item Applying by hand the specialization for invertible
      operators to sumOfProd  --- see 
      section~\ref{subsec:red-inv} --- results in AD overheads
      of $2.4\times$ and $2.8\times$ for $D_1$ and $D_2$,
      which offers good efficiency. 

\item The impact of the Irwim rule (see re-write~\ref{Irwim})
      is highlighted by the case of vectorized multiplication: 
      the AD overhead is under $4\times$ and reasonably close
      to that of multiplication.
      In comparison, PPAD (\textbf{Cmp}) has overhead of
      $797\times$ on $D_2$, because it differentiate such a reduction 
      into scans whose operators are defined on arrays but are not
      vectorized; such reductions/scans are ill supported
      by the Futhark compiler. 
      
\item Matrix multiplication triggers the general-case algorithm
      but results in ``reasonable'' AD overheads between
      $4.6 - 7.6\times$. This are still between 
      $2\times - 5.3\times$ faster than the PPAD algorithm.

\item in all tested cases our algorithm for reduce is faster than
      PPAD --- this is not surprising since our algorithm is 
      AD efficient, while PPAD's piggybacks on the algorithm
      for scans, which is claimed to not be AD efficient.
\end{itemize}


\subsection{Scan}
\label{subsec:eval-scan}

The right-hand side of figure~\ref{fig:eval-red-scan-perf} presents
the performance of scan:
%
%
\begin{itemize}
\item The specialized rule for addition results in small AD
      overheads of under $3\times$, but min and
      multiplication --- which are treated with the general-case
      rule (arity $1$) --- are not far behind, i.e., their 
      AD overheads are under $4.1\times$.

\item The operators defined on tuples of arity $2$, namely
      \fut{linFnComp} and \fut{sumOfProd} still offer decent
      AD overheads of under $5.3\times$, where \fut{linFnComp}
      is more efficient because it benefits from the optimization
      of RBD sparsity, discussed in section~\ref{rbd-sparsity}.

\item The application of Iswim rule (see re-write~\ref{Iswim}) 
      in the case of vectorized multiplication is very beneficial,
      resulting in AD overheads of $0.44\times$, which suggests
      that Futhark compiler should always apply it.
      In comparison, PPAD (\textbf{Cmp}) is $76\times$ slower,
      for the same reasons as the ones discussed for reduce.

\item For matrix multiplication, which benefits from the RBD
      sparsity optimization, the AD overhead reaches a peak of $8.1\times$
      for $3\times 3$ matrices, but then decreases for $4\times 4$ 
      and $5\times 5$ matrices up
      until $6.9\times$, which seems to indicate that
      performance remains a constant factor away from the primal.

\item Our algorithm is faster than PPAD in all evaluated case,
      with speedups typically ranging between $1.3\times - 2.25\times$,
      but we surmise that on operators defined on high-dimensional
      tuples ($d\geq 3$) that do not benefit from the
      sparsity optimizations, PPAD will be significantly
      more efficient than ours.
\end{itemize}


\begin{figure}
\hspace{-2ex}
\begin{minipage}[b]{0.99\columnwidth}
  \small\centering
  \begin{tabular}{l|rr|rr|rr|rr|}
   \textbf{Op} & \multicolumn{2}{c|}{\fut{+}} & \multicolumn{2}{c|}{\fut{map2 (+)}} & \multicolumn{2}{c|}{\fut{ min }} & \multicolumn{2}{c|}{\fut{map2 min}} \\
               & \textbf{Prim} & \textbf{OV} & \textbf{Prim} & \textbf{OV} & \textbf{Prim} & \textbf{OV} & \textbf{Prim} & \textbf{OV} \\\hline
               & \textbf{Gb/s} & $\AdOv{AD}$  & \textbf{Gb/s} & $\AdOv{AD}$  & \textbf{Gb/s} & $\AdOv{AD}$  & \textbf{Gb/s} & $\AdOv{AD}$  \\\hline
  $\mathbf{D_{1,1}}$ & $1176$& $1.5\times$ & $610$  & $1.6\times$ & $1176$ & $1.9\times$ & $621$ & $3.0\times$ \\
  $\mathbf{D_{1,2}}$ & $1087$& $1.4\times$ & $552$  & $1.5\times$ & $1156$ & $2.0\times$ & $557$ & $4.1\times$ \\
  $\mathbf{D_{1,3}}$ & $806$ & $1.4\times$ & $135$  & $1.1\times$ & $324$  & $3.0\times$ & $64$  & $3.2\times$ \\
  $\mathbf{D_{2,1}}$ & $1947$& $1.9\times$ & $975$  & $1.9\times$ & $1932$ & $2.1\times$ & $1041$& $9.6\times$ \\
  $\mathbf{D_{2,2}}$ & $1941$& $1.9\times$ & $950$  & $1.8\times$ & $1932$ & $2.1\times$ & $1009$& $13.7\times$ \\
  $\mathbf{D_{2,3}}$ & $871$ & $1.4\times$ & $164$  & $1.1\times$ & $344$  & $3.9\times$ & $86$  & $4.0\times$ \\  \hline\hline
  \textbf{Op} & \multicolumn{2}{c|}{\fut{*}}& \multicolumn{2}{c|}{\fut{map2 (*)}} & \multicolumn{2}{c|}{\fut{sumOfProd}} & \multicolumn{2}{c|}{\fut{satAdd}} \\\hline
  $\mathbf{D_{1,1}}$ & $1212$ & $2.1\times$ & $643$ & $4.0\times$ & $1495$ & $29.4\times$ & $1136$ & $29.7\times$ \\
  $\mathbf{D_{1,2}}$ & $1111$ & $2.0\times$ & $569$ & $3.9\times$ & $1448$ & $48.6\times$ & $1136$ & $50.6\times$ \\
  $\mathbf{D_{1,3}}$ & $303$  & $2.6\times$ & $64$  & $4.5\times$ & $177$  & $9.4\times$  & $302$  & $21.2\times$ \\
  $\mathbf{D_{2,1}}$ & $1955$ & $2.4\times$ & $1013$& $8.6\times$ & $2348$ & $45.8\times$ & $1937$ & $50.3\times$ \\
  $\mathbf{D_{2,2}}$ & $1945$ & $2.4\times$ & $977$ & $9.0\times$ & $2353$& $84.0\times$ & $1923$ & $92.0\times$ \\
  $\mathbf{D_{2,3}}$ & $265$  & $2.1\times$ & $80$  & $12.4\times$& $141$ & $8.0\times$  & $264$  & $19.5\times$ \\  \hline\hline
  \end{tabular}
\end{minipage}
\vfill
\vspace{-1ex}
\caption{Reverse-AD Performance of \fut{reduce_by_index}.}
\label{fig:eval-red-by-index}
\end{figure}

\subsection{Reduce By Index}
\label{subsec:eval-red-by-ind}

Figure~\ref{fig:eval-red-by-index} shows the performance of
differentiating reduce-by-index. We do not use a competitor since
we are not aware of work on differentiating this construct at a
high-level.   Key observations are:
\begin{itemize}
\item The base cases ($+$,$*$,\fut{min}) are efficiently
      differentiated with overheads under $2.4\times$, except
      for \fut{min} on $D_{1/2,3}$, which corresponds to the
      largest histogram of length $50$K that fits only in global
      memory. The slowdown is due to the lifted operator requiring
      a mutex lock instead of using atomic primitives such as
      \fut{atomicAdd} or \fut{atomicMul} as with the other cases.
      
\item Vectorized operators incurs larger overheads, due to reasons
      similar to the ones discussed for \fut{min} above
      (see also section~\ref{subsubsec:red-by-ind-discussion}).
      However, applying by hand the adaptation of \textsc{Irwim}
      for reduce by index (see re-write~\ref{Irbiwim})
      is very beneficial, resulting in the following column of
      AD overheads:
      $[2.7\times, 2.8\times, 2.9\times, 3.6\times, 3.5\times, 1.7\times]^T$
      for vectorized multiplication (and similar for \fut{min}).
      We are planning to implement this transformation in
      the differentiation pass.

\item \fut{sumOfProd} and \fut{satAdd} are dispatched to the general-case
      algorithm that involves sorting and is inefficient, resulting
      in AD overheads as high as $92\times$. The rationale behind this
      is discussed at the end of section~\ref{subsec:gen-case-red-by-ind}.
      An obvious optimizations would be to improve the underlying
      sorting implementation.

\item Supporting the invertible-operator refinement, discussed in
      section~\ref{subsubsec:inv-red-by-ind} and applied by hand
      to \fut{sumOfProd}, results in very efficient differentiation,
      i.e. the following AD overhead column 
      $[1.8\times, 1.8\times, 1.1\times, 2.1\times, 2.1\times, 1.1\times]^T$.
\end{itemize}

\section{Related Work}
\label{sec:relatedwork}

The most related work is the one of PPAD~\cite{PPAD} that
presents algorithms for high-level reverse-mode differentiation
of reduce and scan. We have compared with it throughout the
paper: Essentially our treatment of reduce is superior,
as ours is AD efficient, and their treatment of scan is
theoretically superior to the one presented in this
paper and also practically superior in the case of
high-dimensional tuples when the RBD sparsity does
not apply.

A body of work has investigated how to differentiate
(parallel) functional array languages at a high level,
i.e., before parallelism is mapped to the hardware. 
Dex~\cite{10.1145/3473593} uses a technique where the
program is first linearized, producing a linear map,
then this linear map is transposed producing
the adjoint code. Dex supports accumulators, which are
discriminated by the type system, intuitively, into parallel
or sequential loops, but does not support second-order 
parallel constructs such as scan and reduce-by-index. 
%

$\widetilde{F}$~\cite{10.1145/3341701} proposes an AD
implementation that is applied to a nested-parallel program
and uses the forward mode, along with rewrite rules for
exploiting sparsity in certain cases.  
DiffSharp~\cite{baydin2015diffsharp} is a library for AD
that aims to make available to the machine learning (ML)
community, in convenient form, a range of AD techniques,
including, among others, nesting of forward/reverse mode AD
operations, efficient linear algebra primitives, and a 
functional API that emphasizes the use of higher-order
functions and composition.

PRAD~\cite{PRAD} is a parallel algorithm for reverse-mode AD of
recursive fork-join programs that (i) is provably work efficient,
(ii) has span within a polylogarithmic factor of the original
program, and (iii) supports Cilk fork-joint parallelism,
without requiring parallel annotations.
Evaluated on $8$ ML applications, PRAD is reported to achieve
$1.5\times$ AD overhead and $8.9\times$ speedup on 18 cores.

None of these approaches propose specific AD algorithms
for differentiating reduce, scan or reduce-by-index.
The algorithms presented in this paper are part of
Futhark's AD system~\cite{futhark-ad-sc22}, that 
supports (forward and) reverse-mode differentiation of
nested-parallel programs. The key difference is that
reverse-AD avoids using tape by a redundant-execution
technique and by techniques that are aimed to rely on
dependence analysis of loops~\cite{SummaryMonot,OanceaSetCongrDynAn}.
The AD implementation benefits from various compiler
optimizations~\cite{futhark-size-types,henriksen2014bounds,futhark-fusion,futhark-mem-sc22,futhark-ppopp,futhark-tuning}
and specialized code generation~\cite{Futhark:redomap,Futhark:segredomap,histo-sc20,futhark-scan}.

Another rich body of work refers to the implementation
of AD algorithms in the imperative context, where 
parallelism is already mapped to hardware, e.g, by
means of low-level APIs such as OpenMP and Cuda.  
Enzyme~\cite{enzymeNeurips} applies AD on low-level compiler
representation, thus taking advantage of both pre- and post-AD
compiler optimizations.  Since the support for AD is built in
the low-level compiler (LLVM), their approach naturally
achieves AD interoperability~\cite{enzyme-mpi-openmp}
across languages, e.g., Julia, and parallel APIs, such
as OpenMP, MPI and Cuda.  In particular, the AD algorithm
for Cuda~\cite{enzyme-gpu} makes use of AD-specific GPU
memory optimizations including caching tape values in
thread-local storage as well as memory-aware adjoint
updates.   However, we speculate that if the target Cuda
kernels is already maxed out in terms of resource usage
then the tape would need to be mapped in global memory,
which will degrade the AD performance.   
This is typically the case for the
kernels generated from reduce(-by-index) and
especially scan~\cite{futhark-scan,futhark-sgm-scan},
which uses a single-pass implementation~\cite{Merrill2016SinglepassPP}.
Reversely, Enzyme's low-level approach of differentiating
memory would likely offer better performance for
cases such as reduce-by-index with non-invertible 
operators (e.g., \fut{satAdd}).
 
In the context of parallel API such as OpenMP and MPI,
other reverse-AD implementations have been proposed,
either by compiler transformations~\cite{huckelheim2021source}
or by overloading techniques~\cite{CoDiPack-1,CoDiPack-2}.

Reverse AD has also been implemented in DSLs for stencil
computations~\cite{stencil-ad}, and tensor 
languages~\cite{bernstein2020differentiating} that
support constrained forms of loops, which do not require
the use of tapes.

ML practitioners use tools such as Tensorflow~\cite{abadi2016tensorflow},
PyTorch~\cite{paszke2019pytorch} or
JAX~\cite{jax2018github,frostig2018compiling} that restrict the
programming interface, but offer well-tuned primitives for AI.
A practically important direction is to promote AD 
interoperability across popular 
languages~\cite{PyTorch-Julia-AD,mixed-lang-AD}, which bears similarities
to prior work on supporting generics in computer algebra~\cite{mapal_synasc}.

Finally, the time-space tradeoff for reverse-mode AD is
systematically studied by Siskind and Pearlmutter~\cite{divide-and-conq},
and Tapenade~\cite{Araya-Polo2004DFA} supports a wealth of
checkpointing techniques. Other approaches aimed at 
sequential code include ADOL-C~\cite{griewank1996algorithm},
and Stalingrad~\cite{lambda-backprop}.

In conclusion, none of the imperative or functional approaches
(other than PPAD) have proposed AD algorithms specific to
reduce, scan and reduce-by-index, or evaluated them for GPU execution. 

\section{Conclusions}

We have presented reverse-mode differentiation algorithms for
reduce, scan and reduce by index second-order parallel array
combinators. Interestingly, the general-case algorithm re-writes
the differentiation of a construct in terms of other,
less-efficient ones: reduce's re-write uses scans, reduce-by-index
uses multi-scan (implemented by sorting), and scan's
re-write is not AD efficient.

However, we have also shown that for most cases of practical
interest, specializations that enable efficient differentiation
are possible: (i) vectorized operators are reduced to scalar
ones and then differentiated, (ii) invertible 
operators allow reduce(-by-index) to be differentiated
in terms of map/reduce-by-index constructs, and (iii) sparsity
optimization allows reasonably-efficient differentiation of scans
with (tuple-based) matrix-multiplication operators, which seem 
to remain a constant-factor away from the primal.  

Most important, we have reported, to our knowledge, the first
evaluation of reverse AD of said constructs in the context
of GPU execution, which constitutes a useful baseline for future
work.

\section*{Acknowledgments}

We would like to acknowledge Troels Henriksen and Robert Schenck
for their invaluable contributions to implementing AD in Futhark.
We credit Troels with the idea of differentiating the classical
work-efficient implementation of scan in the case of 
un-vectorized operators on arrays.
This work has been supported by the UCPH Data+ grant: 
\emph{High-Performance Land Change Assessment} and by the
the Independent Research Fund Denmark (DFF) under the grant
\emph{Monitoring Changes in Big Satellite Data via Massively Parallel AI}.

\bibliographystyle{ACM-Reference-Format}
\bibliography{ad-pbbs.bib}

\newpage

\section{Appendix}
\label{appendix}

For completeness, figure~\ref{fig:ppad-scan} shows the
Futhark implementation of the PPAD rule for reverse
differentiating scan~\cite{PPAD}, which we have used
in our evaluation.

\begin{figure}[t]
\begin{lstlisting}[language=futhark, numbers=left]
def op_bar_1 't (op : t -> t -> t) 
                (x: t, y: t, r_b: t) : t =
  let op' b a = op a b in  vjp (op' y) x r_b

def op_bar_2 't (op : t -> t -> t) 
                (x: t, y: t, r_b: t) : t =
  vjp (op x) y r_b

def op_lft 't (plus: t -> t -> t) (op : t -> t -> t)
              (x1: t, a1: t, y1_h: t)
              (_x2: t, a2: t, y2_h: t) : (t, t, t) =
  let z = plus (op_bar_1 op (x1, a1, y2_h)) y1_h
  in  (x1, op a1 a2, z)

def scan_bar [n] 't (zero: t) (plus: t -> t -> t)
       (op :  t -> t -> t) (e : t) (u : [n]t)
       (x_b : [n]t) : [n]t =
  let x = scan op e u
  let u_lft = map (\i -> if i<n-1 then u[i+1] else e)
                  (iota n)
  let m = zip3 x u_lft x_b
  let (_, _, x_hat) = unzip3 <|
        scan_right (op_lft plus op) (e, e, zero) m
  let x_rht = map (\i -> if i==0 then e else x[i-1])
                  (iota n)
  in  map (op_bar_2 op) (zip3 x_rht u x_hat)
\end{lstlisting}\vspace{-2ex}
\caption{Futhark Implementation of PPAD~\cite{PPAD} Reverse-AD Rule for Scan}
   \label{fig:ppad-scan}
\end{figure}

\end{document}